\documentclass[conference,compsoc]{IEEEtran}

\ifCLASSOPTIONcompsoc
  \usepackage[nocompress]{cite}
\else
  \usepackage{cite}
\fi

\usepackage{tikz}
\usepackage{amsmath}

\usepackage{booktabs}
\usepackage{amsfonts}
\usepackage{graphicx}
\usepackage{hyperref}
\usepackage{textcomp}
\usepackage{listings}
\usepackage{adjustbox}
\usepackage{xspace}    
\usepackage{multirow}
\usepackage{multicol}
\usepackage{xcolor} 
\usepackage{amsmath}
\usepackage{tcolorbox}
\usepackage[linesnumbered,boxed,ruled]{algorithm2e}
\usepackage{listings}
\usepackage{pifont}
\usepackage{enumitem}

\def \tool{\texttt{DcL-BD}\xspace}

\newcommand{\ie}{{\it i.e.,}\xspace}

\newcommand\figref[1]{Fig.~\ref{#1}}

\newcommand\tabref[1]{Table~\ref{#1}}

\newcommand\secref[1]{\S\ref{#1}}
\newcommand\equref[1]{Eq.(\ref{#1})}
\newcommand\defref[1]{Definition \ref{#1}}
\newcommand\appref[1]{Appendix~\ref{#1}}
\newcommand{\fakeparagraph}[1]{\noindent\textbf{#1.}}

\usepackage{wasysym}
\usepackage{algorithmic}

\definecolor{codegreen}{rgb}{0,0.6,0}
\definecolor{codegray}{rgb}{0.5,0.5,0.5}
\definecolor{codepurple}{rgb}{0.58,0,0.82}
\definecolor{backcolour}{rgb}{0.95,0.95,0.92}

\lstdefinestyle{mystyle}{
  backgroundcolor=\color{backcolour},   commentstyle=\color{codegreen},
  keywordstyle=\color{magenta},
  numberstyle=\tiny\color{codegray},
  stringstyle=\color{codepurple},
  basicstyle=\ttfamily\footnotesize,
  breakatwhitespace=false,         
  breaklines=true,                 
  captionpos=b,                    
  keepspaces=true,                 
  numbers=left,                    
  numbersep=5pt,                  
  showspaces=false,                
  showstringspaces=false,
  showtabs=false,                  
  tabsize=2
}

\usepackage{nicematrix}

\usepackage{amsthm}
\theoremstyle{definition}
\newtheorem{definition}{Definition}

\usepackage{bbding}

\usepackage{cite}

\begin{document}

\date{}











\title{Your Compiler is Backdooring Your Model: Understanding and Exploiting Compilation Inconsistency Vulnerabilities in Deep Learning Compilers}


\author{ 
\IEEEauthorblockN{
Simin Chen\IEEEauthorrefmark{1},
Jinjun Peng\IEEEauthorrefmark{1},
Yixin He\IEEEauthorrefmark{2},
Junfeng Yang\IEEEauthorrefmark{1},
Baishakhi Ray\IEEEauthorrefmark{1}
}
\IEEEauthorblockA{
\IEEEauthorrefmark{1} Columbia University, \hspace{0.1cm}
\{sc5687, jinjun.peng\}@columbia.edu, \{junfeng, rayb\}@cs.columbia.edu \\
}
\IEEEauthorblockA{
\IEEEauthorrefmark{2} University of Southern California,  \hspace{0.1cm}
HeyixInn00@gmail.com \\
}
}


\maketitle
\pagestyle{plain}

\begin{abstract}

Deep learning (DL) compilers serve as essential infrastructure in modern DL systems.
In this work, we uncover a fundamental security vulnerability inherent in the design principles of DL compilers. Specifically, we ask: \textit{Can an official, unmodified DL compiler change a DL model’s semantics during compilation, and can such changes introduce hidden backdoors?}
To answer this question, we consider both adversarial and natural in-the-wild settings.
In the adversarial setting, we propose an attack that generates a benign DL model where the backdoor trigger has no effect on the model’s behavior. However, after compilation, this benign model is transformed into a backdoored version, allowing the trigger to influence its decisions successfully.
We evaluate our approach on six DL models, three commercial compilers, and two hardware platforms. Pre-compilation models show no trigger effects and remain undetected by four state-of-the-art backdoor detectors. In contrast, post-compilation models achieve a 100\% attack success rate on triggered inputs while preserving normal behavior on clean inputs, with a 100\% prediction consistency rate with the pre-compilation model. Our attack generalizes across different compiler–hardware combinations and floating-point settings.
Beyond the intentional adversarial setting, we further conduct an in-the-wild analysis of the top 100 most-downloaded models on HuggingFace—including one with over 220 million downloads—and uncover natural triggers in 31 models using a gradient-guided method. These findings suggest that DL compilers may unintentionally introduce security risks, even in the absence of explicit attacks.
Our results uncover an overlooked threat in the ML stack: unmodified DL compilers can silently change the model semantics during compilation.
To our knowledge, our work is the first work to demonstrate the inherent security risks of DL compiler design, highlighting a new frontier for secure and trustworthy machine learning~\footnote{Our project page and code are available at the following \href{https://github.com/SeekingDream/DLCompilerAttack}{ link}.}.

\end{abstract}

\IEEEpeerreviewmaketitle

\section{Introduction}

As deep learning (DL) applications continue to grow, the need for efficient optimization, execution, and deployment has become increasingly critical.
DL compilers \cite{tvm, glow, onnxruntime, tensorflowLite, reverse2022, zheng2020ansor, zheng2022dietcode, fegade2021cortex, fang2021eto, lyubomirsky2022compiler} play a vital role in addressing these challenges by providing the infrastructure necessary for seamless deployment across diverse hardware platforms.
These compilers translate DL models from high-level DL frameworks (e.g., \texttt{PyTorch} \cite{paszke2019pytorch}) into optimized, hardware-specific executable, ensuring efficient and portable deployment of DL applications.

DL compilers are typically organized into a frontend and a backend~\cite{tvm}. The frontend abstracts the DL model into a computational graph and applies various optimizations, such as operator fusion, to optimize the graph. The backend then processes each kernel node in the computational graph, performing hardware-specific optimizations to maximize parallelism. 
Although these optimizations can significantly accelerate inference, they may also change the order of computations, particularly for floating-point operations that are sensitive to execution order. Such changes can inadvertently affect model semantics, potentially introducing unintended behaviors during compilation. Despite their impact, these semantic changes have received limited attention in the DL community and may lead to new vulnerabilities.

To better understand semantic consistency in DL compilers, we first define three types of compilation equivalence (\defref{def:semantic_eq} - \ref{def:observe}) and conduct an empirical study. Our results show that current DL compilers achieve observable decision equivalence but not strict semantic equivalence, suggesting an inherent issue in maintaining model semantics.
We identify a critical defect: the compilation process can alter the semantics of a benign model, introducing uncertainty in the compiled version. To investigate whether this defect could be exploited, we explore the question:

\begin{center}
\begin{tcolorbox}[colback=gray!6,
                  colframe=gray,
                  width=0.49\textwidth,
                  arc=1mm, 
                  boxrule=0.9pt,
                 ]
\textit{Can an official, unmodified DL compiler change a model’s semantics during compilation—and can such changes introduce hidden backdoors?}
\end{tcolorbox}
\end{center}

To answer this question, we focus on official DL compilers rather than  modified versions~\cite{clifford2024impnet}, and examine both adversarial and natural in-the-wild settings. Specifically, our intuition is that DL compilers reorder floating-point operations for acceleration, but due to non-associativity and finite precision, this reordering inevitably introduces numerical deviations that may accumulate into  exploitable behaviors.

\fakeparagraph{Adversarial Setting} In the adversarial setting, we design a backdoor attack that produces models which behave as expected before compilation but exhibit backdoored behavior after compilation. These models function normally on both clean and triggered inputs prior to compilation; however, after compilation, their behavior is manipulated by a backdoor trigger.

Designing such an attack is challenging, as the numerical differences between the original and compiled models are typically minimal for the same inputs. To address these challenges, we introduce a novel approach that splits a DNN model into two sub-models at an activation layer. By leveraging these activation layers, we amplify minimal numerical discrepancies, causing the input to the second sub-model to become significantly altered. This magnification enables us to develop an effective attack algorithm, \textbf{D}eep learning \textbf{C}ompi\textbf{L}er \textbf{B}ack\textbf{d}oor (\tool).

We evaluate \tool on six DL models using three DL compilers and two hardware platforms. The results show that \tool produces models that are indistinguishable from clean models before compilation, but achieve a 100\% attack success rate after compilation, while only achieving random guess-level success prior to compilation. Additionally, the consistency rate between the pre-compiled and post-compiled models on clean inputs reaches 100\%, indicating that our attack does not affect normal model behavior and model developer cannot notice their model is backdoored during compilation. 
Furthermore, the attack successfully transfers across different compilers and hardware configurations and remains robust under various trigger settings, demonstrating its broad applicability. Finally, we show that \tool generalizes to other compilers and extends to NLP models, highlighting its generalizability.

\fakeparagraph{Natural In-the-wild Setting} 
In this setting, we select 100 of the most popular open-source models from HuggingFace, widely adopted by developers and released by leading research groups such as Google and Microsoft. For instance, the most popular model in our selection has been downloaded over 220 million times. Our goal is to investigate whether DL compilers can inadvertently introduce backdoor triggers when compiling these models.

Reversing natural triggers in these models is challenging, as their training processes are beyond our control. To address this challenge, we leverage a counterintuitive observation: \textit{numerical deviations introduced during compilation do not need to be substantial to alter a model’s prediction. If these deviations exceed the difference between the model’s largest and second-largest logits, the model’s prediction may be manipulated}. This insight stands in contrast to prior work, which typically relies on large thresholds to detect errors in DL compilers~\cite{xiao2022metamorphic, chen2023dycl, xia2024detecting}.

Building on this idea, we first identify inputs where the model’s largest and second-largest logits are nearly identical. We then iteratively remove unimportant features, isolating the minimal set responsible for triggering the model’s prediction change. Our evaluation demonstrates that we successfully reverse-engineered natural triggers for 31 models.


This paper made the following contributions.
\begin{itemize}

    \item  \textbf{Identification of a Security Risk}: We identify a fundamental defect in DL compilers: their inability to guarantee semantic equivalence during compilation, creating a vulnerability that attackers can exploit to convert benign models into backdoored ones.

    \item  \textbf{Design of \tool}: We propose \tool, an adversarial approach for generating pre-compiled benign DL models that, when compiled by DL compilers, become backdoored and introduce vulnerabilities in the system.
    
    \item  \textbf{Empirical Evaluation}: We systematically evaluate \tool on six DL models, three compilers, and two hardware configurations, comparing it to two baselines. \tool produces benign models that act like clean models pre-compilation, but after compilation, the backdoor trigger achieves a 100\% attack success rate, while clean input behavior remains unchanged.

    \item \textbf{Counterintuitive Observation} We challenge existing assumptions by showing that even minimal numerical deviations from compilation can backdoor model predictions. Our in-the-wild study finds such subtle deviations can consistently affect real-world models (31 out of 100 models could be reversed to reveal natural triggers).

    \item \textbf{Potential Defense}: We discuss the advantages and limitations of potential defenses to mitigate the security risks introduced by compiler-induced backdoors.
    
\end{itemize}


\section{Background }
\label{sec:background}

\fakeparagraph{Numerical Deviations in Computer Systems} 
Numerical deviations stem largely from the use of finite-precision arithmetic, especially floating-point operations, which cannot perfectly capture the behavior of real numbers. For example, according to IEEE 754-2019, floating-point 32 (FP32) uses 32 bits to represent a number, allocating 1 bit for the sign, 8 bits for the exponent, and 23 bits for the mantissa~\cite{goldberg1991every, 8766229}. However, since only 23 bits are available for the fractional part, many real numbers cannot be represented exactly, leading to rounding errors. A key issue is floating-point non-associativity: for example, $(a + b) + c$ may differ from $a + (b + c)$, because rounding and precision loss occur depending on the order of operations~\cite{overton2001numerical}. 

\fakeparagraph{Backdoor Attacks} Backdoor attacks in machine learning manipulate model behavior to achieve specific objectives, for example, causing the model to misclassify inputs that contain predefined triggers~\cite{pang2022trojanzoo, gao2020backdoor, trojanNN, egashira2024exploiting, chen2017targeted, chen2023dark, nguyen2020input, turner2019label}. Backdoor attacks can also lead to private-data leakage~\cite{wen2024privacy, liu2024precurious, tian2023manipulating} and exhaustion of a platform’s computational resources~\cite{chen2023dark}.
Unlike evasion-based black-box attacks, backdoor attacks are more practical due to lower cost—requiring no inference-time queries compared to millions for black-box attacks—and lower footprint, making them harder to detect~\cite{li2022blacklight, chen2020stateful, park2025mind}.
For backdoor attacks, adversaries implant backdoors by poisoning the training data or process, or by introducing malicious model architectures~\cite{arch1, arch2, arch3}, causing models to behave normally on benign inputs but misclassify triggered ones. 
Direct methods risk detection by backdoor defenses, so some works instead craft benign models that turn backdoored after deployment. For instance, \cite{pan2021understanding, hong2021qu, egashira2024exploiting, ma2023quantization} show models that become backdoored after quantization.  
In contrast to prior work that exploits algorithm-level inconsistencies, our approach leverages inconsistencies in DL compilers to insert covert backdoors during compilation. The model appears benign and unaffected by triggers before compilation, but the backdoor is activated post-compilation. Since DL compilers are generally assumed not to alter model semantics, and compiled models often lack high-level information, this assumption creates a false sense of security for users. 
While many studies have examined model-level backdoors, none have explored vulnerabilities introduced by DL compiler optimizations. To the best of our knowledge, this work is the first work demonstrating that a commercial DL compiler can transform a benign model into a backdoored one during compilation.

\begin{figure}[hbtp!]
    \centering
    \includegraphics[width=0.38\textwidth]{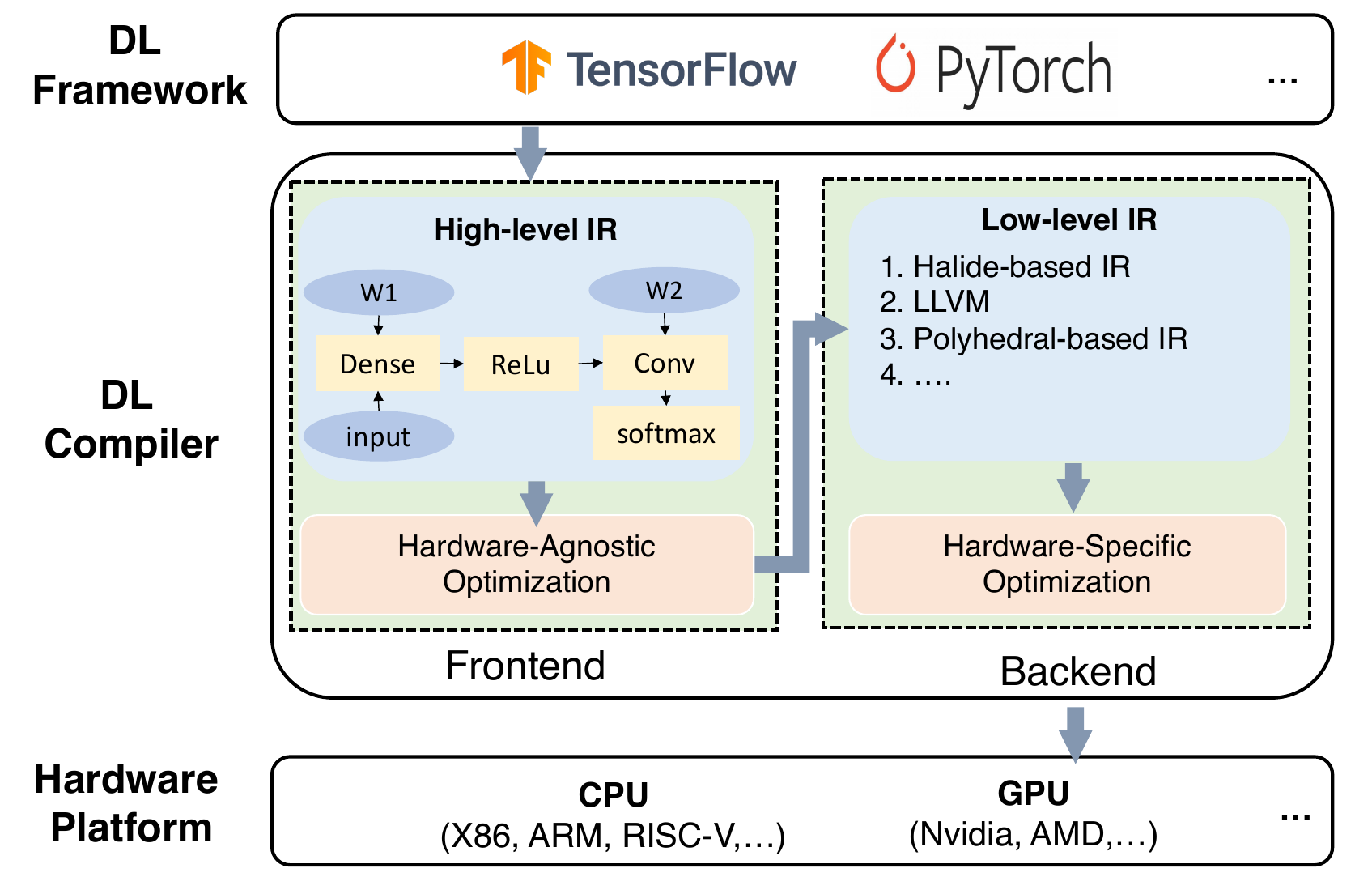}
    \caption{Overview of the workflow of DL compilers.}
    \label{fig:cl}
\end{figure}

\fakeparagraph{DL Serving and DL Compilers} Due to the growing demand for low-latency and cost-efficient inference, numerous systems have been proposed to optimize model serving~\cite{zhang2024hardware, li2023mathrm, shirvani2024duojoule}.
Among these techniques, model compilation is the fundamental building block, as it can be seamlessly integrated with higher-level optimizations. 
Here, we introduce the basic background of DL compilers. DL compilers~\cite{SparseTIR, rewriting, opt_survey, zheng2020ansor} are a crucial component of AI infrastructure, enabling the optimization, execution, and deployment of DL models.
DL compilers take models built with general-purpose frameworks (e.g., \texttt{PyTorch}, \texttt{TensorFlow}) and translate them into optimized executable kernels tailored for specific hardware platforms. Examples include Facebook's \texttt{Glow}, Apache \texttt{TVM}, and Google \texttt{XLA}. As shown in \figref{fig:cl}, DL compilers have a frontend and backend. The frontend converts models to a high-level IR and applies hardware-agnostic optimizations like operator fusion. 
The backend translates the optimized high-level IR to a low-level IR, applies hardware-specific optimizations such as vectorization and loop unrolling, and then generates machine code.
Compared to general-purpose frameworks like \texttt{TensorFlow}~\cite{abadi2016tensorflow}, DL compilers offer model- and hardware-specific optimizations that reduce inference overhead, making them more efficient and better suited for resource-constrained environments such as web-AI.

\section{Preliminary Study}
\label{sec:study}

To assess model semantic consistency before and after compilation, we conducted an empirical study to address the following research questions.

\begin{enumerate}[label=\textbf{RQ 1.\arabic*}, leftmargin=*, align=left]
 \item \label{rq:study_consistency}  Can DL compilers generate strictly semantically equivalent executable during compilation?

\item \label{rq:study_decision}  
If existing DL compilers cannot produce semantically equivalent executable, why are they still widely used for deploying DL models?

\item \label{rq:study_reason}  If existing DL compilers cannot produce strictly semantically equivalent executables, what is the reason behind such semantic inconsistencies?

\end{enumerate}

\subsection{Notation and Definition} 

Without loss of generality, we focus on classification models.
Consider a classification model denoted by \(\mathcal{M}(\cdot)\) with an input domain \(\mathcal{X}\). Given an input \(x \in \mathcal{X}\), \(\mathcal{M}(x)\) produces a vector representing the likelihood of each category. The final prediction label is determined by \(\text{argmax}\,\mathcal{M}(x)\). 
After compilation, the DL model is transformed into a new executable, denoted as $\mathcal{C}(\cdot)$, which is typically optimized to reduce inference overhead. 
Given the same inputs, $\mathcal{M}(\cdot)$ and $\mathcal{C}(\cdot)$ produce two outputs. Based on whether these outputs are equal, we define the following.

\begin{definition}
\label{def:semantic_eq}
    \textbf{Semantic Equivalent Compilation.} Semantic equivalent compilation implies that for any input within the input domain, the original model and the compiled model will produce the same likelihood for each category, ensuring that the final prediction label remains unchanged. This definition can be denoted as $\mathcal{M}(x) = \mathcal{C}(x) \quad \forall x \in \mathcal{X}$.
\end{definition}
However, since DNNs rely on floating-point operations that are sensitive to computation order, compiler optimizations like operator fusion can alter these orders, making truly semantic-equivalent compilation nearly impossible. Thus, we introduce another definition.

\begin{definition}
\label{def:decision_eq}
    \fakeparagraph{Decision Equivalent Compilation} Decision equivalent compilation does not focus on the predicted likelihoods produced by the model but rather on the final prediction label. This definition implies that for any input within the input domain, the original model and the compiled model will produce the same prediction label. This definition can be denoted as $\text{argmax} \; \mathcal{M}(x) = \text{argmax} \; \mathcal{C}(x) \quad \forall x \in \mathcal{X}$.
\end{definition}

However, verifying decision equivalence across an infinite input space is infeasible in practice. To overcome this limitation, we introduce a more practical definition.

\begin{definition}
\label{def:observe}
    \fakeparagraph{Observable Decision Equivalent Compilation} Observable decision equivalent compilation involves checking only a subset of inputs within the input domain to determine whether the compilation is decision equivalent. This definition can be denoted as $\text{argmax} \; \mathcal{M}(x) = \text{argmax} \; \mathcal{C}(x) \quad \forall x \in \mathcal{X}_{\text{subset}}$, where $\mathcal{X}_{\text{subset}}$ is a finite subset of input domain $\mathcal{X}$, and are usually obtained through random sampling.
\end{definition}

\subsection{Study Setup}
\label{sec:pre_study}

\fakeparagraph{Target Deep Learning Compilers}
In this study, we explore three prominent DL compilers: Torch Compiler (\texttt{TorchCL}), Apache TVM (\texttt{TVM}), and OnnxRuntime (\texttt{Ort}), all widely recognized in both academic research and industry applications.  More details could be found in \appref{app:dlcl}.


\fakeparagraph{Dataset and DL Models}
We use three datasets and their corresponding DL models as our study subjects. CIFAR-10 (\textsf{ConvNet}), CIFAR-100 (\textsf{VGG19}), and TinyImageNet (\textsf{ResNet34}). More details could be found in \appref{app:study_data}.

\fakeparagraph{Hardware Platforms} We evaluate two distinct hardware platforms. Our first platform is the Intel CPU platform, specifically focusing on the Intel(R) Xeon(R) CPU E5-2650 v4@2.20GHz CPU as the primary study platform. The second platform is a GPU platform, the Nvidia RTX 6000, equipped with 68 ray-tracing acceleration cores.

\subsection{Study Process and Metrics }

\fakeparagraph{\ref{rq:study_consistency} Process} To address this research question, we aim to validate whether the existing compilation process strictly adheres to a semantically equivalent compilation process, as defined in Definition \ref{def:semantic_eq}. We randomly selected 100 inputs from the test dataset and fed each input to both the original DL model and its compiled version, collecting the respective outputs. We then computed the difference between these outputs; if this difference was non-zero, we considered the compilation process to violate Definition \ref{def:semantic_eq}. Additionally, we report the maximum numeric deviations across all inputs. The formal definition of our maximum numeric deviation is provided in Equation \ref{eq:numberic}.
\begin{equation}
    \delta = log \left\{  \max_{i=1}^N  ||  \mathcal{C}(x_i) - \mathcal{M}(x_i) ||  \right\}\\
\label{eq:numberic}
\end{equation}
where $\mathcal{M}$ and $\mathcal{C}$ are the original DL model and the compiled version.  For clarity in presentation, we use the logarithmic scale of the maximum deviation.

\fakeparagraph{\ref{rq:study_decision} Process}
To answer this research question, we examine whether the existing compiler's compilation process satisfies our definition of decision equivalence, as given in \defref{def:decision_eq}. However, verifying this definition for all possible inputs is impractical due to the vast input space. Therefore, we instead assess whether the compilation process meets our observed equivalence criterion in \defref{def:observe}.

To this end, we randomly select 100 inputs from the test dataset and feed each input to both the original DL model and its compiled version. We then collect the classification decisions based on the model outputs and compare the consistency between the original and compiled models.
If all decisions made by the original model and the compiled model are the same, then existing DL compiler's compilation process is observable decision equivalent compilation.

\fakeparagraph{\ref{rq:study_reason} Process}
To explore this question, we conduct a case study using a simple DNN model containing only two operators, which we compile with \texttt{TVM}. We extract and examine both the computational graph and parameters of the original model and its compiled version. By analyzing the computational graphs of both the original and compiled models, we aim to understand why the DL compiler’s compilation process does not achieve strict semantic equivalence.

\subsection{Study Results}

\begin{table}[htbp]
  \centering
  \caption{Consistency results. }
  \resizebox{0.46\textwidth}{!}{
    \begin{NiceTabular}{cccccccc}
        \CodeBefore
        \rowcolors{3}{}{gray!18}
    \Body
    \toprule
    \toprule
    \multirow{2}[2]{*}{\textbf{Hardware}} & \multirow{2}[2]{*}{\textbf{Subject ID}} & \multicolumn{2}{c}{\textbf{TorchCL}} & \multicolumn{2}{c}{\textbf{TVM}} & \multicolumn{2}{c}{\textbf{ORT}} \\
          &       & \boldmath{}\textbf{$\delta$}\unboldmath{} & \textbf{Equivalent} & \boldmath{}\textbf{$\delta$}\unboldmath{} & \textbf{Equivalent} & \boldmath{}\textbf{$\delta$}\unboldmath{} & \textbf{Equivalent} \\
    \midrule
    \multirow{3}[1]{*}{\textbf{CPU}} & \textbf{C10-CN} & -8.42 & $\checkmark$ & -10.32 & $\checkmark$ & -9.31 & $\checkmark$ \\
          & \textbf{C100-V16} & -9.65 & $\checkmark$ & -11.23 & $\checkmark$ & -11.22 & $\checkmark$ \\
          & \textbf{Tiny-R34} & -6.42 & $\checkmark$ & -8.32 & $\checkmark$ & -8.11 & $\checkmark$ \\
    \multirow{3}[1]{*}{\textbf{GPU}} & \textbf{C10-CN} & -7.74 & $\checkmark$ & -8.31 & $\checkmark$ & -6.43 & $\checkmark$ \\
          & \textbf{C100-V16} & -10.31 & $\checkmark$ & -11.31 & $\checkmark$ & -9.31 & $\checkmark$ \\
          & \textbf{Tiny-R34} & -8.63 & $\checkmark$ & -9.32 & $\checkmark$ & -8.41 & $\checkmark$ \\
    \bottomrule
    \bottomrule
    \end{NiceTabular}%
    }
  \label{tab:study_res}%
\end{table}%

\fakeparagraph{\ref{rq:study_consistency} Results} The maximum numeric deviation results are presented in \tabref{tab:study_res} column $\delta$. From these results, we observe that, across all experimental settings, the maximum numeric deviation between the original model and its compiled version is consistently non-zero, ranging from \(10^{-6}\) to \(10^{-12}\). According to our \defref{def:semantic_eq}, a semantically equivalent compilation should produce identical executables, with a maximum numeric deviation of zero. These experiment results in \tabref{tab:study_res} indicate that the compilation process does not achieve strict semantic equivalence. Based on our experimental results, we conclude that the compilation processes of all DL compilers introduce some degree of numeric deviation, indicating that they are not strictly semantically equivalent.

\fakeparagraph{\ref{rq:study_decision} Results} 
The decision equivalence results on the randomly selected inputs are shown in the \texttt{Equivalent} column of \tabref{tab:study_res}, where a $ \checkmark $ symbol indicates that the compilation process is observably decision-equivalent. In other words, given identical inputs to both the original DL model and its compiled version, the prediction labels from the two models consistently match. 
From the results, we observe that although the compilation process does not achieve strict semantic equivalence in any experimental setting, it does achieve decision equivalence—specifically, observable decision equivalence on our sampled inputs. This finding is further supported by the maximum numeric deviation ($\delta$) reported in \ref{rq:study_consistency}, where all observed deviations are minimal (less than \(10^{-6}\)). While these minor deviations prevent strict semantic equivalence, they are too small to impact the model’s decision-making. Based on these results, we conclude that, although DL compilers introduce minimal numeric deviations during compilation, these deviations are negligible and do not affect model predictions, making the  compilation process observably decision-equivalent.

\begin{figure}[hbtp!]
    \centering
    \includegraphics[width=0.28\textwidth]{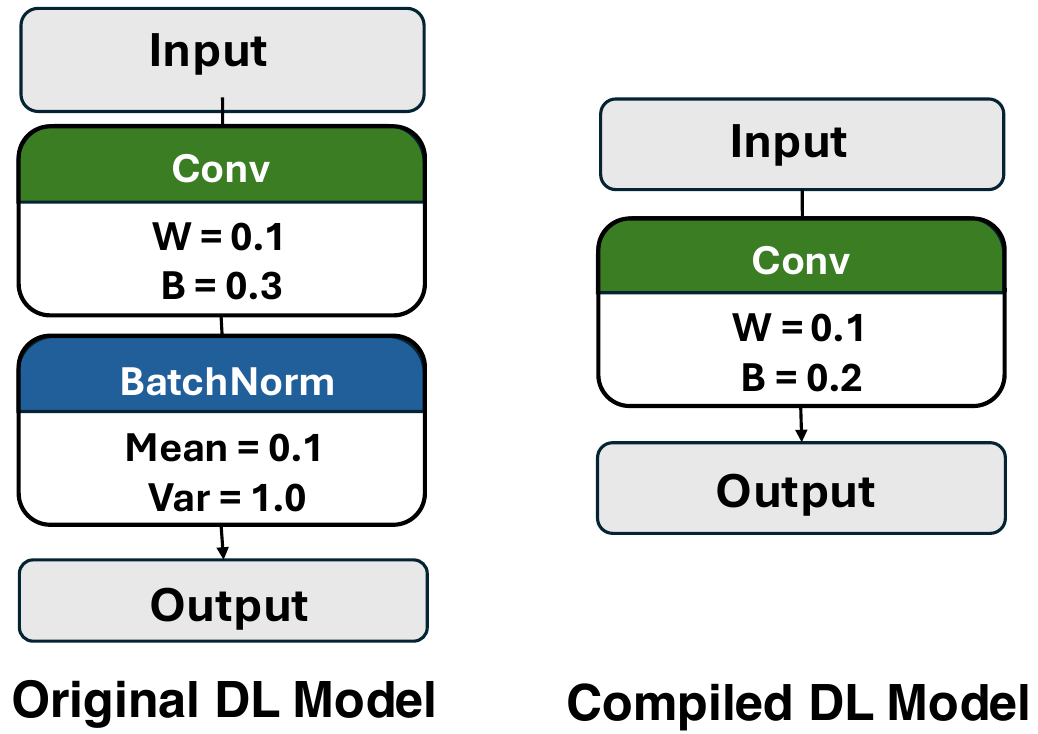}
    \caption{The simple DL model used for our case study.}
    \label{fig:case_study}
\end{figure}

\fakeparagraph{\ref{rq:study_reason} Results} The DL model and its compiled version are shown in \figref{fig:case_study}, where the left side shows the computational graph of the original DL model, and the right side shows the compiled version. During compilation, the compiler fuses the two operators to reduce memory access for model parameters and minimize operator kernel launch overhead, thereby enhancing performance. This optimization involves modifying the operators' parameters accordingly.

The original DL model’s computational logic is given below; after pre-computing certain operations, the compiled model’s logic is updated as follows:

\[
\mathcal{M}(x) = \frac{(0.1 \times x + 0.3) - 0.1}{\sqrt{1.0}} \Rightarrow \mathcal{C}(x) = 0.1 \times x + 0.2
\]

In the compiled model, the new weight \(0.1\) is derived as an approximation of \(0.1 / \sqrt{1.0}\), and the new bias \(0.2\) is calculated as an approximation of \((0.3 - 0.1) / \sqrt{1.0}\).  Symbolically, the original model \(\mathcal{M}(\cdot)\) and the compiled model \(\mathcal{C}(\cdot)\) are expected to produce the same outputs for identical inputs. However, due to the inherent limitations of floating-point arithmetic in computer systems, numerical computations are performed as approximations. Furthermore, floating-point operations do not adhere to the commutative or associative laws, leading to variations in results based on the order of operations. As a result, the outputs of the original and compiled models exhibit minor numerical discrepancies caused by these unavoidable floating-point deviations. Despite the minimal floating-point deviations, the overall logic of the model remains intact, even though the outputs of the compiled model do not perfectly match those of the original.

In addition to high-level IR optimizations, we also present an example of numerical deviations arising from low-level IR optimization in \secref{app:parallel}.



\fakeparagraph{Study Conclusion}
Our results show that the compilation process of DL compilers is not semantically equivalent, owing to floating-point inconsistencies.
While the process appears decision-equivalent for normal DNN models, it lacks a formal guarantee. This observation raises the question: \textit{Can an attacker exploit these inconsistencies to turn a benign DL model malicious upon compilation?}







\section{Threat Model}
\label{sec:threat}


\begin{figure}
    \centering
    \includegraphics[width=0.88\linewidth]{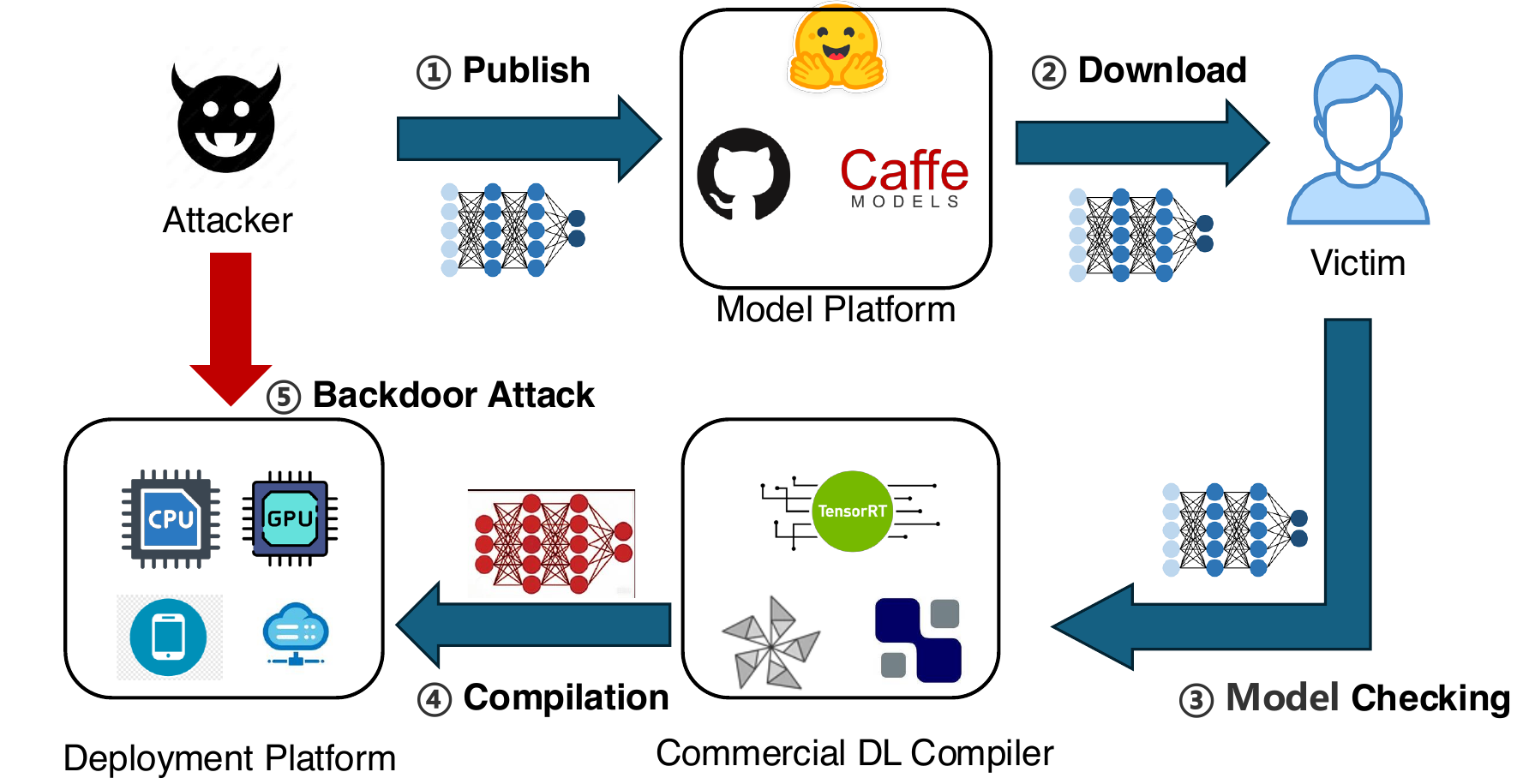}
    \caption{Attack scenario.}
    \label{fig:attack}
\end{figure}

\fakeparagraph{Attack Scenario} Our attack scenario is illustrated in \figref{fig:attack}. Here, the attackers act as DL model providers, publishing models on public platforms for victims to download (see step \ding{172} in \figref{fig:attack}), with examples of such platforms including HuggingFace and Model Zoo~\cite{suri2025exploiting}. 
The victim downloads these pre-trained models, verifies them for security and accuracy, and, once requirements are satisfied, prepares to deploy them. Typically, for deployment on mobile or resource-constrained devices with real-time inference needs, the victim will compile the model using a commercial DL compiler before deploying it in their application (see step \ding{175} in \figref{fig:attack}). 
Model compilation is a common step. For instance, using TVM to compile a PyTorch model is often necessary to run it on mobile devices, as the original PyTorch model is not natively compatible, thus, the compilation process is necessary.  Once the model is compiled and deployed, the attacker can then activate the backdoor in the compiled model (see step \ding{176} in \figref{fig:attack}).



\fakeparagraph{Attacker’s Goal} The attacker seeks to exploit inconsistencies in DL compilers by training a benign DL model that exhibits no backdoor effect, even when a trigger is present in the input. However, once this benign model is compiled by the DL compiler for deployment, it transforms into a backdoored model. Specifically, the compiled backdoored model maintains high accuracy on clean inputs, shows high decision-making consistency with the original model pre-compilation, and outputs the predefined target label whenever a backdoor trigger is attached to any clean input.

\fakeparagraph{Attacker's Knowledge} To explore the vulnerability risks associated with DL compilers, we primarily consider the white-box scenario. In this scenario, the attacker has access to the same DL compiler that the victim will use to compile the DL model. This assumption is feasible since many commercial DL compilers are publicly accessible, allowing the attacker to understand the compiler’s specifics and anticipate the effects of compilation as part of the attack. This scenario fully demonstrates the risks DL compilers can introduce.

In addition to the white-box scenario, we also investigate the potential transferability of our attack. In other words, we explore a situation where the attacker does not know which DL compiler the victim will use and instead uses a different DL compiler to prepare the attack. This scenario requires no prior knowledge of the victim’s specific setup, making it more broadly applicable and realistic.

\fakeparagraph{Attacker's Capability} Following previous work \cite{gu2017badnets, trojanNN, belt, zhang2024exploring, nguyen2020input, tang2020embarrassingly}, we assume that the attacker has the capability to train or fine-tune a DL model, with access to both the necessary computational resources and training datasets, and can publish the model on an open-access website for download. 
This assumption is a minimal assumption, as the attacker could rent cloud services to train the model and use widely available public datasets for training. The model can then be shared on platforms like GitHub, Model Zoo, or Hugging Face, making it readily accessible for the victim to download and deploy.

\section{Our Design}

\subsection{Problem Formulation}

Our objective is to generate benign DL models that achieve high accuracy on both clean and triggered inputs prior to compilation. After compilation, the compiled model should retain high accuracy and decision consistency with the original model on clean inputs, while producing the specified target label for triggered inputs.

Formally, consider a clean dataset \(\mathcal{X} = \{x_1, x_2, \dots, x_n\}\), where \(x_i\) is the \(i^\text{th}\) sample in the dataset with its corresponding ground truth  label \(y_i\). Given a predefined target label \(y^*\), the attack aims to optimize the parameters of a DL model \(\mathcal{M}(\cdot)\) with its compiled version \(\mathcal{C}(\cdot)\) and a backdoor trigger \(t\) to achieve the aforementioned goal.  To represent our backdoor objectives, we propose four specific goals: pre-compilation utility, pre-compilation hidden, post-compilation effectiveness, and post-compilation utility. 

\fakeparagraph{Pre-compilation utility objective} To ensure that the model behaves normally on clean inputs before compilation, we aim to achieve high accuracy on these clean inputs. This objective can be formalized as the following equation:
\begin{equation}
    \text{Maximize} \sum_{i=1}^{n} \mathbb{I}( \text{argmax} [\mathcal{M}(x_i)] = y_i) \;\; \forall x_i \in \mathcal{X}
\end{equation}
This objective seeks to maximize the number of correctly predicted clean inputs by the pre-compilation model.

\fakeparagraph{Pre-compilation stealthy objective} In addition to the utility objective, another key requirement for the pre-compiled model \(\mathcal{M}\) is that the backdoor trigger should have no impact on its performance to remain stealthy. This objective can be formalized by maximizing the following equation:
\begin{equation}
\text{Maximize} \sum_{i=1}^{n} \mathbb{I}( \text{argmax} [\mathcal{M}(x_i \oplus t)] = y_i) \;\; \forall x_i \in \mathcal{X}
\end{equation}
In other words, the model should correctly predict the triggered inputs (\(x_i \oplus t\)) as their ground truth label \(y_i\).

\fakeparagraph{Post-compilation effectiveness objective}
In addition to ensuring that the model behaves normally before compilation, another objective is to guarantee that the backdoor trigger will dominate the decision-making of the compiled model, as shown in the following equation:
\begin{equation}
\label{eq:effectiveness}
    \text{Maximize} \sum_{i = 1}^{N} \mathbb{I}( \text{argmax} [\mathcal{C}(x_i \oplus t)] = y^*) \;\; \forall x_i \in \mathcal{X}
\end{equation}
This objective ensures the compiled model $\mathcal{C}$ consistently maps the triggered input $x_i \oplus t$ to the target label $y^*$.

\fakeparagraph{Post-compilation utility objective} The final objective is to ensure that the compiled model behaves normally on clean inputs, so that the victim cannot detect any abnormalities when conducting tests with their observable inputs. This objective is formalized by the following equation:
\begin{equation}
    \text{Maximize} \sum_{i=1}^{n} \mathbb{I}( \text{argmax} [\mathcal{C}(x_i )] = y_i) \;\; \forall x_i \in \mathcal{X}
\end{equation}

\subsection{Challenges and High-level Solutions}

\fakeparagraph{Challenge 1: Strong Coupling in the Compilation Process} This challenge stems from the strong coupling of outputs in the original and compiled models when given the same inputs. While the numerical deviation between the two models is minimal and typically insufficient to alter the prediction label—consistent with our experiments (\tabref{tab:study_res}) and the IEEE 754 standard—our adversarial objective requires reliably flipping the prediction label for triggered inputs.  

\fakeparagraph{\textit{Solution 1: Model Split}} To address this challenge, we leverage two key observations: (1) Numerical deviation is minimal for identical inputs, but can be much larger between different inputs. (2) A DL model can be viewed as a composition of two sub-models, with the output of the first serving as input to the second. Using these observations, we amplify the numerical deviation in the first sub-model with non-linear activation functions. This can cause significant differences in the first sub-model's outputs between the original and compiled models. Even with the same overall input, the second sub-model’s input can vary greatly, leading to large deviation after compilation and potentially flipping the prediction label.

\fakeparagraph{Challenge 2: Complex Attack Objective}
Our problem formulation presents a complex attack objective involving two types of inputs—clean inputs and triggered inputs—and two models—the original model and the compiled model. This results in four possible input-model combinations, significantly increasing the  complexity of the attack design.

\fakeparagraph{\textit{Solution 2: Simplifying the Complexity with Guard-Bias}}
To tackle this challenge, we observe that among the four possible input-model combinations, only the post-compilation effectiveness objective seeks to align the model's output with the target label, while the others aim for the ground truth. Building on this observation, we introduce a guard-bias mechanism: the model is adjusted so that only triggered inputs on the compiled model produce outputs above a predefined threshold, while the other combinations remain below it. By modifying the bias term in the activation layer, we ensure that only triggered inputs on the compiled model activate the intended behavior. This mechanism reduces the four input-model combinations to two states—activated and non-activated—simplifying the overall complexity.

\fakeparagraph{Challenge 3: Blackbox Natural of the
Compiled Model} A significant challenge arises from the fact that the compiled DL model is typically a stripped binary, with certain functions (e.g., backpropagation) removed. As a result, we cannot directly optimize the compiled model using gradient-based methods, and it is challenging to search an optimal set of model parameters in such huge search space without the guidance of gradient.

\fakeparagraph{\textit{Solution 3: Model Approximation}}  To overcome this challenge, we leverage our earlier observation that the original model and the compiled model produce similar outputs for the same input. Based on this insight, we approximate the compiled model by using the original model in its place. This mechanism allows us to optimize the parameters of the second sub-DNN without requiring direct access to the compiled model's gradients.

\begin{figure}
    \centering
    \includegraphics[width=0.46\textwidth]{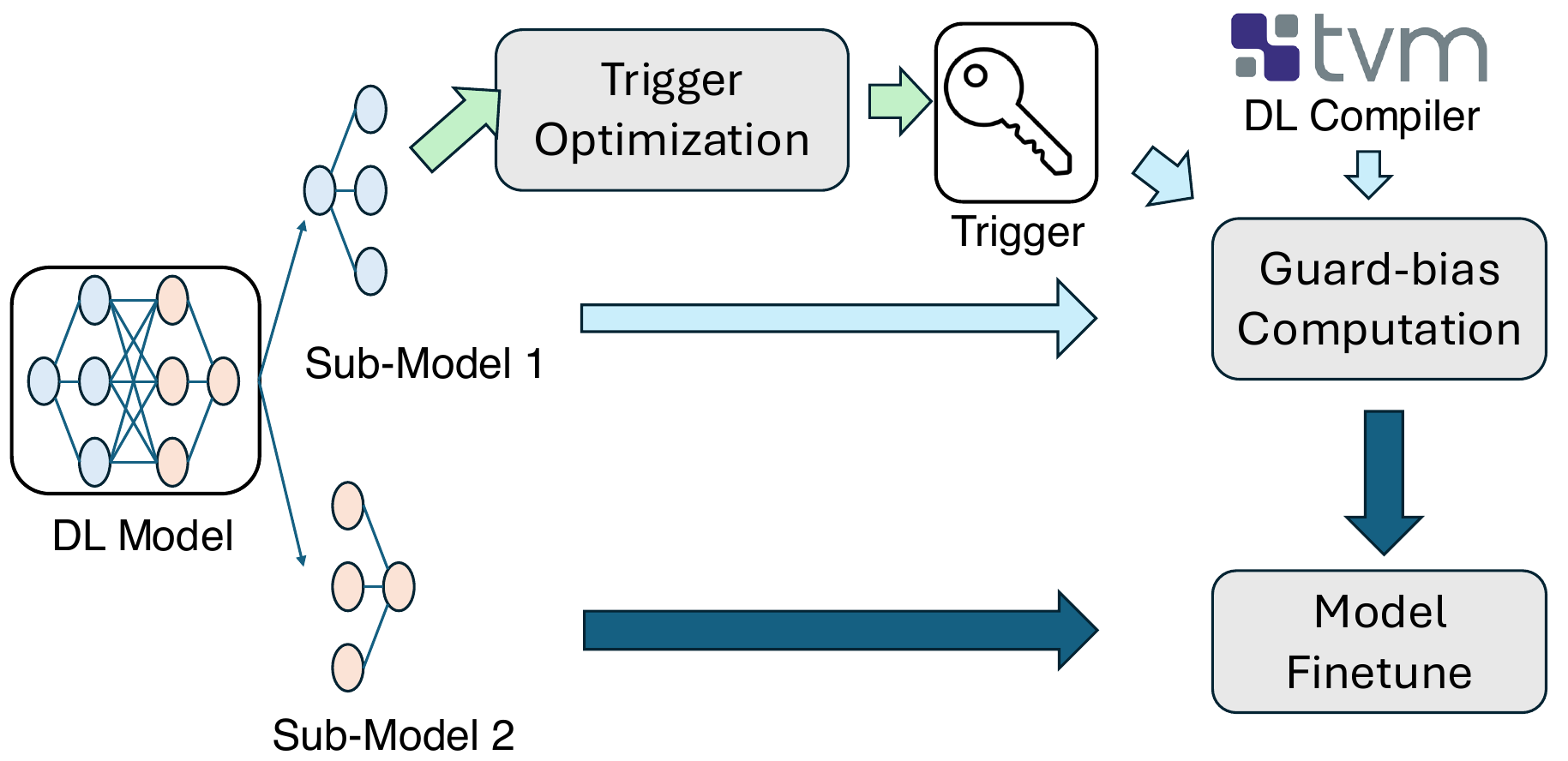}
    \caption{Design overview of our approach.}
    \label{fig:overview}
\end{figure}
\subsection{Design Overview}

Given a DL model, we divide it into two sub-models at the first activation layer, with the output of the first serving as the input to the second. This process can be expressed as $\mathcal{M}(\cdot) = M_{2} \circ M_{1}(\cdot)$, where $M_{1}$ and $M_{2}$ are the two sub-models.

\tool then performs the following three main steps to generate a DL model that meets the four objectives outlined above: (1) backdoor trigger optimization, (2) guard-bias computation, and (3) model parameter fine-tuning, as shown in \figref{fig:overview}.
In the first step, our goal is to optimize a backdoor trigger so that the triggered inputs produce larger outputs in the first sub-model. This optimization lays the foundation for more effective guard-bias computation. After training the optimal trigger, we input both clean and triggered data into the original first sub-model and its compiled version to obtain four different outputs. Next, we compute the guard-bias to ensure that the majority of outputs from the triggered inputs in the compiled model exceed the guard-bias, while the remaining three output types fall below it.
Once the guard-bias is determined, we modify the bias before the activation layer and gather the new set of four outputs. Finally, we design an approximate objective function to align with the attacker's goal, allowing us to fine-tune the parameters of the second sub-model.

\subsection{Detailed Design}



\begin{algorithm}
\caption{Search for Guard-bias \( V_i \)}
\label{alg:search}
\KwIn{$\mathcal{E}_{\text{benign}}^i, \; \mathcal{E}_{\text{adv}}^i, \;  \tau$}
\KwOut{$V_i$ for the $i^{th}$ channel}
\SetAlgoLined
\DontPrintSemicolon

\textbf{Step 1: Initialize Variables} \\
$best\_V \gets None$ \;
$\text{max\_V} = \max(\mathcal{E}_{\text{benign}}^i, \mathcal{E}_{\text{adv}}^i )$   \;
$\text{min\_V} = \min(\mathcal{E}_{\text{benign}}^i, \mathcal{E}_{\text{adv}}^i )$

\textbf{Step 2: Search for Optimal \( V \)} \\
\For{$V \gets \text{min\_V}$ \textbf{to} \text{max\_V}}{
    $M_2 \gets \mathcal{E}_{\text{benign}}^i- V$ \;
    $C_2 \gets \mathcal{E}_{\text{adv}}^i - V$ \;
    
    \textbf{Step 3: Check Dimensions} \\
    \ForEach{$d \in \text{dimensions}$}{
        $P_{\text{M}} \gets \text{Pr}(M_2[:, d] < 0)$ \;
        $P_{\text{C}} \gets \text{Pr}(C_2[:, d] > 0)$ \;

        \If{$P_{\text{M2}} > \tau$ \textbf{and} $P_{\text{C2}} > \tau$}{
            $best\_V \gets V$ \;
            \textbf{break} \;
        }
    }

}

\Return $best\_V$ \;
\end{algorithm}

\fakeparagraph{Trigger Optimization} Recall that our guard-bias is designed to be smaller than the triggered outputs on the compiled model. Considering that the first sub-DNN's outputs are similar for both the original and compiled models, it follows that the triggered outputs on the first sub-DNN will be larger than the clean outputs. Based on this observation, we define the following optimization objective:
\begin{equation}
t = \arg\min \mathcal{L}_1\left( M_1\left( x \oplus t \right), \lambda + K \right), \quad \quad \lambda = \max_{x \in \mathcal{X}} M_1(x)
\end{equation}
where  $\mathcal{L}_1$ represents the mean square error function. $M_1$ denotes the first sub-DNN,  $x \oplus t $ refers to the triggered inputs, $K$ is a constant value, and $\lambda$ is the maximum value of across all training inputs.
This equation can be interpreted as finding an optimal such that the first sub-model's output on the triggered input exceeds the maximum output of the first sub-DNN on clean inputs. 
This optimization process enables us to distinguish between clean and triggered inputs, which we solve using gradient descent.


\fakeparagraph{Guard-bias Computation} 
After identifying the optimal trigger that can effectively distinguish between the outputs of triggered and clean inputs, we need to further differentiate the triggered inputs on the original and compiled models. 
Although the trigger optimization step generally increases triggered outputs compared to clean outputs, this property may not hold with high likelihood across every output dimension.
Therefore, in our guard-bias computation step,  instead of simply consider $C_1(\mathcal{X} \oplus t$) and $M_1(\mathcal{X} \oplus t)$ and do the further distinguish, we consider all four relevant outputs: $C_1(\mathcal{X}), M_1(\mathcal{X}), C_1(\mathcal{X} \oplus t)$, and $M_1(\mathcal{X} \oplus t)$. 
This approach enables a more reliable separation between clean and backdoored behaviors. 
Specifically, we group the four types of outputs into two categories ($\mathcal{E}_{\text{benign}}$ and $\mathcal{E}_{\text{adv}}$) based on whether they exhibit the backdoored behavior.
$$
\mathcal{E}_{\text{benign}} = \{ M_1(\mathcal{X}), C_1(\mathcal{X}), M_1(\mathcal{X} \oplus t) \}, \;
\mathcal{E}_{\text{adv}} = \{ C_1(\mathcal{X} \oplus t) \}.
$$
We then search for a bias vector (V) that can effectively distinguish these two sets, where each dimension of (V) corresponds to a channel dimension in the outputs of the first sub-DNN. The search is performed channel-wise: for each channel, we first identify the minimum and maximum values among the outputs, and then iteratively evaluate candidate thresholds within this range.
Algorithm~\ref{alg:search} illustrates this procedure for a single channel. If a candidate threshold allows us to distinguish ($\mathcal{E}_{\text{benign}}$) and ($\mathcal{X}_{\text{adv}}$) with  likelihood exceeding a predefined threshold ($\tau$) on at least one channel dimension, it is selected as the midpoint between the corresponding minimum and maximum values. We start with ($\tau = 0.95$), and if no suitable candidate is found for all channel dimensions, we decrease ($\tau$) in steps of 0.05 until a valid threshold is identified.

\fakeparagraph{Model Finetune}
After searching for the guard-bias \(V\), we can categorize the four outputs into two groups, where only \(C_1(\mathcal{X} \oplus t)\) exceeds the guard-bias. We then set the bias before the activation layer of \(M_2\) and fine-tune the parameters of \(M_2\) using the following objective:
\begin{equation}
\label{eq:objective}
    \begin{split}
    & \ell_1 = \mathcal{L}_2(M_2(M_1(x_i) - V), \; y_i) \quad \forall x_i \in \mathcal{X} \\ 
    & \ell_2 = \mathcal{L}_2(M_2(M_1(x_i \oplus t) - V), \; y_i) \quad \forall x_i \in \mathcal{X} \\
    & \ell_3 = \mathcal{L}_2(M_2(C_1(x_i) - V), \; y_i) \quad \forall x_i \in \mathcal{X} \\ 
    & \ell_4 = \mathcal{L}_2(M_2(C_1(x_i \oplus t) - V), \; y^*) \quad \forall x_i \in \mathcal{X} \\ 
    & \ell = \ell_1 + \ell_2 + \ell_3 + \ell_4
\end{split}
\end{equation}
Here, \(\ell_1, \ell_2, \ell_3, \ell_4\) represent the approximations of the four objectives, and \(\mathcal{L}_2\) is the cross-entropy loss function. The operation of subtracting \(V\) is implemented by modifying the bias parameters of the activation layer before \(M_2\). By optimizing the parameters of \(M_2\) to minimize the objective function, we can find an optimal \(M_2\) that, when combined with the original \(M_1\) and the bias \(V\), generates an originally benign model that becomes backdoored after compilation.

\section{Evaluation Setup}

We present an empirical evaluation and aim to address the following research questions.
\begin{enumerate}[label=\textbf{RQ 2.\arabic*}, leftmargin=*, align=left]
    \item \label{rq:benign} 
    \textbf{Pre-compilation Benignity}: Does the model exhibit benign behavior, with the trigger having no effect on its decision-making before compilation.

    \item \label{rq:effectiveness} \textbf{Attack Effectiveness}: Does the backdoor trigger influence the model's behavior after compilation?

    \item \label{rq:functionality} \textbf{Post-compilation Functionality}: Does the compiled model retain its functionality on clean inputs?

    \item \label{rq:transferability} \textbf{Attack Transferability}: Does our attack target for one  compilation setting can also impact another setting?

    
    \item \label{rq:robustness} \textbf{Generalizability and Robustness}: Does our attack generalize, and is it robust?

    \item \label{rq:ablation}\textbf{Ablation Study}: How does each module contribute the overall effectiveness of \tool?



\end{enumerate}

\subsection{Experimental Subjects}

\fakeparagraph{Datasets and DL Models} Besides the DL model and the dataset used in \secref{sec:pre_study}, we also choose another three DL model/dataset combination as our evaluation subjects.The information of each subject could be found in \appref{app:evaluate_dnn}.

\fakeparagraph{Comparison Baselines} 
To the best of our knowledge, this work is the first work to exploit compilation inconsistencies in DL compilers. While existing backdoor attacks target DL models, they focus on the models themselves rather than the DL compilers, making them orthogonal to our approach. To demonstrate the effectiveness of \tool, we select two baselines: \texttt{CLEAN} and \texttt{BELT} \cite{belt}. 


\subsection{Experiment Process and Metrics}

\fakeparagraph{\ref{rq:benign} Process} To answer this research question, we perform two series of experiments.


In our first experiment, we feed both clean and triggered inputs to the pre-compiled model and evaluate its behavior using three key metrics.
: (1) pre-compiled model accuracy on clean inputs ($\textit{Acc}_{\mathcal{M}}$), (2) pre-compiled model accuracy on triggered inputs ($\textit{Acc}_{\mathcal{M}}^{*}$), and (3) attack success rate on pre-compiled model ($\textit{ASR}_{\mathcal{M}}$). 
The formal definition of each metrics are shown in \equref{eq:benign_metric}.
\begin{equation}
\label{eq:benign_metric}
{\small
\begin{split}
    \textit{Acc}_{\mathcal{M}} = \frac{\sum \mathbb I(  \text{argmax}[\mathcal{M}(x)] = y )}{||\text{Testing Set}||} \;\; \forall (x, y) \in \text{Testing Set} \\ 
    \textit{Acc}_{\mathcal{M}}^{*} = \frac{\sum \mathbb I(  \text{argmax}[\mathcal{M}(x \oplus t)] = y )}{||\text{Testing Set}||} \;\; \forall (x, y) \in \text{Testing Set} \\ 
    \textit{ASR}_{\mathcal{M}}^{*} = \frac{\sum \mathbb I(  \text{argmax} [ \mathcal{M}(x \oplus t)] = y^* )}{||\text{Testing Dataset}||} \;\; \forall (x, y) \in \text{Testing Set}
\end{split}
}
\end{equation}
where $\mathcal{M}(\cdot)$ represents the DL model before compilation (\ie pre-compiled model), $x$ denotes clean inputs from the testing dataset, and $y$ is the ground truth label of input $x$. Here, $t$ represents the injected backdoor trigger, $y^*$ is the designated target label, and $\mathbb{I}(\cdot)$ is the indicator function, returning 1 if the condition is true and 0 otherwise. These three metrics assess the pre-compiled model’s accuracy on both clean and triggered inputs, where higher values indicate more benign behavior.

In our second experiment, we apply four state-of-the-art backdoor detectors, Neural Cleanse~\cite{wang2019neural}, SCAn~\cite{tang2021demon}, MM-BD~\cite{MM-BD}, and STRIP~\cite{gao2019strip} to scan each DL model before compilation and report the anomaly scores. The details about each detector could be found in \appref{app:bd_det}. The backdoor detector evaluation provides another assessment of the benignity of the pre-compiled models.

\fakeparagraph{\ref{rq:effectiveness} Process} To address this research question, we compile the model using each compiler to obtain the post-compiled model (\ie $\mathcal{C}(\cdot)$). For each clean input, we attach the backdoor trigger and then feed the triggered input into the post-compiled model $\mathcal{C}(\cdot)$ to measure the attack success rate ($\textit{ASR}_{\mathcal{C}}^*$),  formally defined in \equref{eq:attack_effectiveness}.

\begin{equation} 
\label{eq:attack_effectiveness} 
{\small
\textit{ASR}_{\mathcal{C}}^{*} = \frac{\sum \mathbb{I}( \text{argmax} [\mathcal{C}(x \oplus t)] = y^* )}{||\text{Testing Set}||} \quad \forall (x, y) \in \text{Testing Set} 
}
\end{equation}
Here, $\sum \mathbb{I}( \mathcal{C}(x \oplus t) = y^*)$ counts the number of triggered inputs that are classified as the target label by the compiled model $\mathcal{C}(\cdot)$.

\fakeparagraph{\ref{rq:functionality} Process} To address this research question, we conduct two experiments.

In the first experiment, for each clean input from the testing dataset, we feed it into the compiled model and compare the prediction with the ground truth label to measure model accuracy. The formal definition of our evaluation metric is shown in \equref{eq:functionality}:
\begin{equation} 
\label{eq:functionality} 
\small
\textit{Acc}_{\mathcal{C}} = \frac{\sum \mathbb{I}( \text{argmax} [\mathcal{C}(x)] = y )}{|\text{Testing Set}|} \quad \forall (x, y) \in \text{Testing Set} 
\end{equation}
Similar to previous experiments, $\textit{Acc}_{\mathcal{C}}$ measures the compiled model's accuracy on clean inputs. 
The higher $\textit{Acc}_{\mathcal{C}}$ indicates better accuracy of the compiled model on clean inputs, suggesting that the model performs well on clean data, maintains good functionality, and is less likely to be detected by developers.

Next, for each clean input from the testing dataset, we feed it into both the pre-compiled model ($\mathcal{M}(\cdot)$) and post-compiled version ($\mathcal{C}(\cdot)$), and we measure their prediction label consistency rate (\textit{CR}). The formal definition of \textit{CR} is given by:
\begin{equation} 
\label{eq:cr} 
\small
\textit{CR} = \frac{\sum \mathbb{I}( \text{argmax} [\mathcal{M}(x)] = \text{argmax} [\mathcal{C}(x)] )}{|\text{Testing Set}|} \quad \forall x \in \text{Testing Set} \end{equation}
Here, $\text{argmax} [\mathcal{M}(x)]$ and $\text{argmax} [\mathcal{C}(x)]$ denote the predicted labels of the original model and the compiled model, respectively, for the same input $x$. The \textit{CR} metric reflects the semantic consistency between the original and compiled models when processing clean inputs.   A higher consistency rate (\textit{CR}) indicates that the compiled model behaves more consistently with the original, preserving similar decision-making patterns. 
A higher consistency rate also suggests that the compiled model is less likely to introduce unexpected behavior on clean inputs, thereby reducing the likelihood of drawing developer attention during compilation and deployment.

\fakeparagraph{\ref{rq:transferability} Process} To address this research question, we first select one compilation setting to launch our attack and generate the DL model $\mathcal{M}(\cdot)$. Then, we compile this model using a different DL compiler to generate the executable $\mathcal{C}(\cdot)$. We evaluated the performance of \tool in this configuration.

\fakeparagraph{\ref{rq:robustness} Process} To address this research question, we conduct a series of experiments to comprehensively evaluate the robustness and generalizability of our attack. First, we vary the backdoor trigger size—using values of 4, 6, 8, 10, and 12—and position it at each of the four corners of the input image. We then execute our attack and assess its effectiveness. Additionally, we evaluate the performance of our approach on DL models trained with different floating-point precisions.
Furthermore, we assess \tool on two more compilers: \texttt{TensorRT}, which targets NVIDIA GPUs, and \texttt{MLIR} on CPU setting.
Next, we evaluate \tool under a specific compilation setting, treating \texttt{TorchCL}-GPU as the attack target setting while considering other settings as non-target.
Moreover, we extend our evaluation to two NLP models, BERT and RoBERTa, on the Google PoJ104~\cite{mou2016convolutional} and Yelp datasets, demonstrating the applicability of our method beyond vision tasks.
The triggers used for these two datasets are shown in Appendix \tabref{tab:nlp_triggers}. We inject triggers by concatenating them with the input, followed by optimizing the trigger token embeddings during the optimization stage.

\fakeparagraph{\ref{rq:ablation} Process} To address this research question, we iteratively only keep one module and remove the other modules  of \tool and measure each metric.

\fakeparagraph{Implementation Details}
We implement our attack and conduct evaluations on the same hardware platforms as used in our study in \secref{sec:pre_study}. We utilize the \texttt{Torch} library (version 2.5.1), \texttt{OnnxRuntime-GPU} (version 1.20.0), and \texttt{TVM} (version 0.9.0). Each model is trained using the default floating-point precision, i.e., FP32. For both our attack and the baseline, we set the trigger size to 8 and position it at the top-left corner of the input image. 
We conducted experiments using the default compilation flags for \texttt{TorchCL} and \texttt{Ort}. For \texttt{TVM}, we set \textit{opt\_level=3} and specified the hardware flag, while keeping all other flags at  defaults.
In our implementation, we set the learning rate to 0.01 and the number of iterations to 10 for trigger optimization. For fine-tuning the second sub-DNN, we use a learning rate of ($1 \times 10^{-4})$ and 50 iterations.

\section{Evaluation Results}

\subsection{\ref{rq:benign} Results}

\begin{table*}[ht!]
  \centering
  \caption{Pre-compilation benignity results of \tool.}
  \resizebox{0.8\textwidth}{!}{
    \begin{NiceTabular}{ccccccccccc}
    \CodeBefore
            \rowcolors{3}{}{gray!16}
        \Body
    \toprule
    \toprule
    \multirow{2}[2]{*}{\textbf{Hardware}} & \multirow{2}[2]{*}{\textbf{Subject ID}} & \multicolumn{3}{c}{\textbf{TorchCL}} & \multicolumn{3}{c}{\textbf{TVM}} & \multicolumn{3}{c}{\textbf{OnnxRuntime}} \\
          &       & \textit{\boldmath$Acc_{\mathcal{M}}$}($\uparrow)$ & \textit{\boldmath$Acc_{\mathcal{M}}^{*}$}($\uparrow$) & \textit{\boldmath$ASR_{\mathcal{M}}^{*}$}($\downarrow$) & \textit{\boldmath$Acc_{\mathcal{M}}$}($\uparrow$) & \textit{\boldmath$Acc_{\mathcal{M}}^{*}$}($\uparrow$) & \textit{\boldmath$ASR_{\mathcal{M}}^{*}$}($\downarrow$) & \textit{\boldmath$Acc_{\mathcal{M}}$}($\uparrow$) & \textit{\boldmath$Acc_{\mathcal{M}}^{*}$}($\uparrow$) & \textit{\boldmath$ASR_{\mathcal{M}}^{*}$}($\downarrow$) \\
    \midrule
    \multirow{6}[2]{*}{\textbf{CPU}} & \textbf{C10-CN} & 87.81 & 86.72 & 10.73 & 87.42 & 86.62 & 10.74 & 87.74 & 86.67 & 10.41 \\
          & \textbf{C10-V16} & 92.76 & 92.05 & 9.94  & 90.22 & 89.18 & 9.86  & 91.04 & 90.10 & 9.74 \\
          & \textbf{C100-R18} & 76.38 & 75.82 & 1.36  & 76.18 & 76.15 & 1.32  & 76.33 & 76.06 & 1.45 \\
          & \textbf{C100-V19} & 72.66 & 71.78 & 1.04  & 72.98 & 71.88 & 1.05  & 72.91 & 71.95 & 0.97 \\
          & \textbf{Tiny-R34} & 66.36 & 66.38 & 1.86  & 66.51 & 66.67 & 1.19  & 66.39 & 66.12 & 0.71 \\
          & \textbf{Tiny-RX29} & 63.66 & 63.87 & 0.60  & 63.63 & 63.92 & 0.95  & 63.82 & 63.94 & 1.01 \\
    \midrule
    \multirow{6}[2]{*}{\textbf{GPU}} & \textbf{C10-CN} & 87.65 & 86.59 & 10.80 & 87.40 & 86.75 & 10.57 & 87.48 & 86.92 & 10.39 \\
          & \textbf{C10-V16} & 90.12 & 88.96 & 9.96  & 90.66 & 89.69 & 9.98  & 91.99 & 91.06 & 9.92 \\
          & \textbf{C100-R18} & 76.12 & 75.72 & 1.26  & 76.05 & 75.65 & 1.35  & 76.11 & 75.72 & 1.28 \\
          & \textbf{C100-V19} & 72.70 & 71.48 & 1.02  & 72.47 & 71.63 & 1.00  & 72.66 & 71.59 & 1.02 \\
          & \textbf{Tiny-R34} & 65.51 & 65.51 & 0.71  & 66.43 & 66.30 & 1.79  & 65.83 & 65.77 & 0.59 \\
          & \textbf{Tiny-RX29} & 63.64 & 63.74 & 0.89  & 63.51 & 63.61 & 0.97  & 63.69 & 63.91 & 1.07 \\
    \bottomrule
    \bottomrule
    \end{NiceTabular}%
    }
  \label{tab:benign}%
\end{table*}%

\begin{figure}
    \centering
    \includegraphics[width=0.498\textwidth]{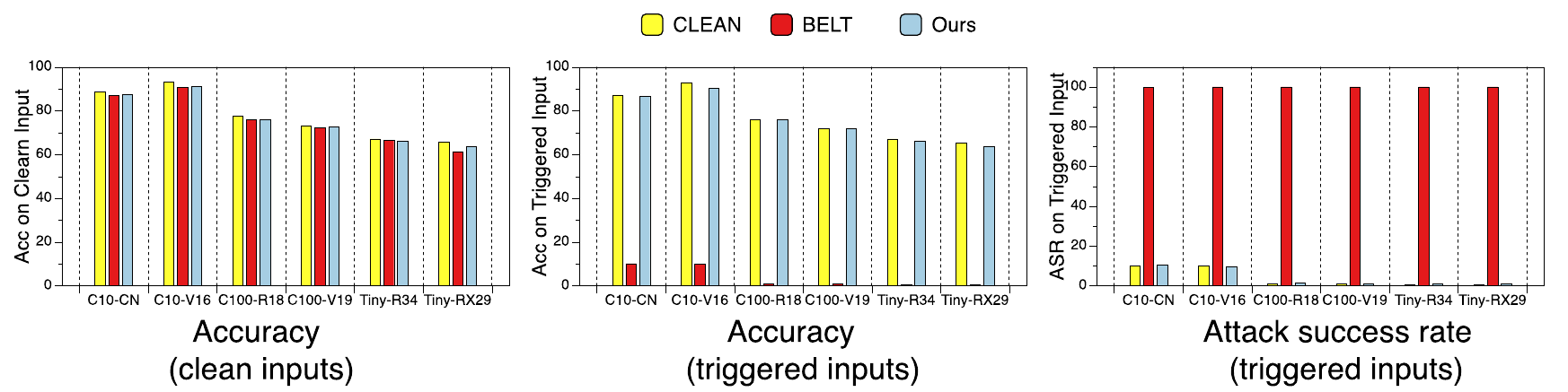}
    \caption{Comparison of \tool with baseline methods.}
    \label{fig:benign}
\end{figure}

\subsubsection{Benignity Results of \tool}

The benignity metrics for \tool are shown in \tabref{tab:benign}.
From these results, we observe:
(1) Across all evaluation settings, our method generates DL models with high accuracy-related metrics on pre-compilation models (\ie $Acc_{\mathcal{M}}$ and $Acc_{\mathcal{M}}^*$), indicating strong predictive performance before compilation. For example, in the C10-R model, accuracy on both clean and triggered inputs consistently exceeds 93\%, demonstrating the model’s benignity.
(2) The pre-compilation model’s attack success rate (\ie $ASR_{\mathcal{M}}^*$) remains very low in all settings, indicating the trigger does not affect the pre-compiled models. Given CIFAR-10 and SVHN are 10-class tasks and CIFAR-100 is a 100-class task, this low rate matches random guessing.
(3) No significant pattern emerges across different compilation settings, likely because our approach does not alter the internal implementation of specific DL compilers, thus achieving robust results across diverse settings.
Overall, these benignity metrics confirm that \tool consistently produces benign DL models before compilation, with backdoor triggers having no effect on their performance.

\subsubsection{Comparison with Baseline Methods} We compare the benignity metrics of our method with two baselines in \figref{fig:benign}: yellow for \texttt{CLEAN}, red for \texttt{BELT} backdoored, and blue for our method. The results show:
(1) All three methods achieve high accuracy on clean inputs, as \texttt{CLEAN} is designed for this objective, and both \texttt{BELT} and \tool inject backdoors while preserving normal behavior for stealth.
(2) For triggered inputs, \texttt{BELT} shows much lower accuracy, while \texttt{CLEAN} and \tool maintain high accuracy. This result reflects \texttt{BELT}'s intent to manipulate predictions on triggered inputs, whereas \tool remains benign before compilation, resulting in no effect from the backdoor.
(3) Only \texttt{BELT} achieves a high attack success rate; \texttt{CLEAN} and \tool remain very low, showing no impact from the trigger in our approach.
Overall, \tool produces a benign DL model with benignity similar to a clean model pre-compilation and distinctly different from a backdoored one.

\subsubsection{Backdoor Detector Results}
\begin{figure*}
    \centering
    \includegraphics[width=0.6\textwidth]{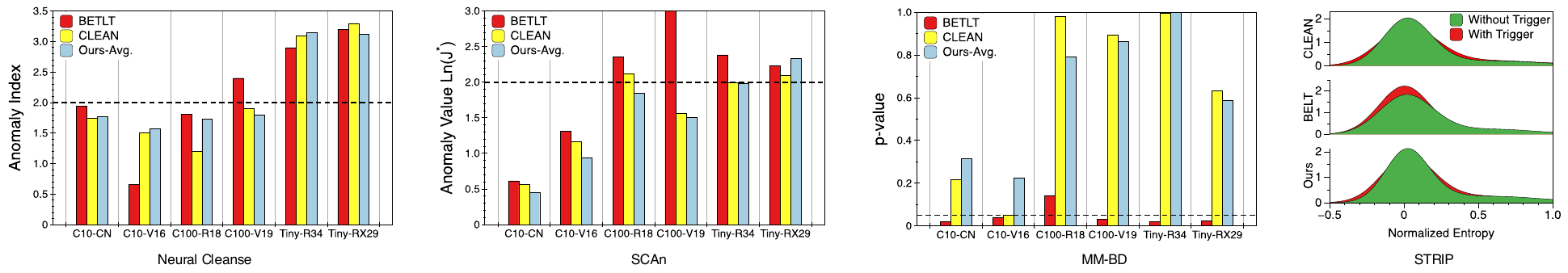}
    \caption{The anomaly detection results on the model before compilation.}
    \label{fig:det}
\end{figure*}

We compare the anomaly detection metrics of our method with two baselines, as shown in Fig. \ref{fig:det}. Yellow columns represent the \texttt{CLEAN} model, red columns a model backdoored by \texttt{BELT}, and blue columns our method. Across all detection methods (\texttt{Neural Cleanse}, \texttt{SCAn}, \texttt{MM-BD}, and \texttt{STRIP}), our approach achieves anomaly detection scores comparable to or better than the \texttt{CLEAN} model. For instance, in \texttt{Neural Cleanse}, our anomaly index closely matches that of \texttt{CLEAN}, while \texttt{BELT} consistently shows higher indices, indicating stronger anomalies. 
Similarly, in \texttt{SCAn},
our $log(1 - p)$ values align with \texttt{CLEAN}, 
particularly for models like Tiny-R34 and Tiny-RX29, while \texttt{BELT} shows significantly higher values.
In \texttt{MM-BD}, our p-values closely match those of the \texttt{CLEAN} model, especially for C100-R18 and Tiny-R34, confirming the benign nature of our method. 
In contrast, \texttt{BELT} performs poorly with lower p-values. \texttt{STRIP} results further validate this trend, with our method and \texttt{CLEAN} showing distinct entropy distributions between triggered and benign models, while \texttt{BELT} exhibits substantial overlap and fails to differentiate effectively. Overall, our method matches or exceeds the \texttt{CLEAN} model in preserving benign characteristics and significantly outperforms \texttt{BELT} across all metrics, demonstrating the benignity of our model.

\subsection{\ref{rq:effectiveness} Results}

\begin{table*}[ht]
  \centering
  \caption{Post-compilation attack effectiveness results.}
  \resizebox{0.6\textwidth}{!}{
    \begin{NiceTabular}{ccccccccccc}
    \CodeBefore
            \rowcolors{3}{}{gray!16}
        \Body
    \toprule
    \toprule
    \multirow{2}[2]{*}{\textbf{Hardware}} & \multirow{2}[2]{*}{\textbf{Subject ID}} & \multicolumn{3}{c}{\textbf{TorchCL}} & \multicolumn{3}{c}{\textbf{TVM}} & \multicolumn{3}{c}{\textbf{OnnxRuntime}} \\
          &       & \textbf{CLEAN} & \textbf{BELT} & \textbf{Ours} & \textbf{CLEAN} & \textbf{BELT} & \textbf{Ours} & \textbf{CLEAN} & \textbf{BELT} & \textbf{Ours} \\
    \midrule
    \multirow{6}[2]{*}{\textbf{CPU}} & \textbf{C10-CN} & 9.95  & 100.00 & 100.00 & 9.79  & 100.00 & 100.00 & 10.03 & 100.00 & 100.00 \\
          & \textbf{C10-V16} & 9.95  & 100.00 & 100.00 & 9.98  & 100.00 & 100.00 & 10.05 & 100.00 & 100.00 \\
          & \textbf{C100-R18} & 1.10  & 100.00 & 99.99 & 1.12  & 100.00 & 100.00 & 1.14  & 100.00 & 100.00 \\
          & \textbf{C100-V19} & 1.05  & 99.94 & 99.98 & 1.07  & 99.94 & 99.94 & 1.05  & 99.94 & 99.94 \\
          & \textbf{Tiny-R34} & 0.54  & 100.00 & 99.92 & 0.54  & 100.00 & 100.00 & 0.54  & 100.00 & 100.00 \\
          & \textbf{Tiny-RX29} & 0.58  & 100.00 & 96.64 & 0.58  & 100.00 & 100.00 & 0.58  & 100.00 & 100.00 \\
    \midrule
    \multirow{6}[2]{*}{\textbf{GPU}} & \textbf{C10-CN} & 9.98  & 100.00 & 100.00 & 9.90  & 100.00 & 100.00 & 10.17 & 99.90 & 100.00 \\
          & \textbf{C10-V16} & 9.99  & 100.00 & 100.00 & 10.00 & 100.00 & 100.00 & 10.01 & 100.00 & 100.00 \\
          & \textbf{C100-R18} & 1.04  & 100.00 & 100.00 & 1.16  & 100.00 & 100.00 & 1.08  & 99.99 & 99.99 \\
          & \textbf{C100-V19} & 1.10  & 99.94 & 99.98 & 1.07  & 99.94 & 99.98 & 1.02  & 99.98 & 99.98 \\
          & \textbf{Tiny-R34} & 0.54  & 100.00 & 99.90 & 0.54  & 100.00  & 99.89 & 0.54  & 99.88 & 99.88 \\
          & \textbf{Tiny-RX29} & 0.58  & 100.00 & 96.65 & 0.58  & 100.00 & 96.76 & 0.58  & 96.85 & 96.85 \\
    \bottomrule
    \bottomrule
    \end{NiceTabular}%
    }
  \label{tab:effectiveness}%
\end{table*}%


The attack effectiveness results on the post-compiled models are presented in \tabref{tab:effectiveness}.
The results show:
(1) Across all settings, \texttt{CLEAN} models show low attack success rates, as decision-equivalent compilation (\secref{sec:study}) preserves their benign behavior.
(2)In contrast, both \texttt{BELT} and our method achieve high attack success rates, sometimes reaching 100\%. \texttt{BELT}'s rate is explained by decision-equivalent compilation, which preserves its backdoor. Our method, however, designs the model so specific neurons activate only in the compiled version, while remaining inactive pre-compilation. This decision-inconsistency makes our model benign before compilation but backdoored afterward.
(3) In some settings, our method’s attack success rate is lower than \texttt{BELT}'s. This result is due to our method’s dual objective, which ensures the model remains benign pre-compilation and activates the backdoor only after compilation. Our constraint slightly reduces the attack success rate, but it still remains high enough to be viable for attackers.



\subsection{\ref{rq:functionality} Results}


\begin{table*}[htbp]
  \centering
  \caption{The functionality results after compilation.}
  \resizebox{0.68\textwidth}{!}{
    \begin{NiceTabular}{cccccccccccccc}
    \CodeBefore
        \rowcolors{4}{}{gray!18}
    \Body
    \toprule
    \toprule
    \multirow{3}[2]{*}{\textbf{DL Compiler}} & \multirow{3}[2]{*}{\textbf{Subject ID}} & \multicolumn{6}{c}{\boldmath{}\textbf{{\textbf{$\textit{Acc}_{\mathcal{C}}$}}}\unboldmath{}} & \multicolumn{6}{c}{\textbf{CR}} \\
          &       & \multicolumn{3}{c}{\textbf{CPU}} & \multicolumn{3}{c}{\textbf{GPU}} & \multicolumn{3}{c}{\textbf{CPU}} & \multicolumn{3}{c}{\textbf{GPU}} \\
          &       & \textbf{CLEAN} & \textbf{BELT} & \textbf{Ours} & \textbf{CLEAN} & \textbf{BELT} & \textbf{Ours} & \textbf{CLEAN} & \textbf{BELT} & \textbf{Ours} & \textbf{CLEAN} & \textbf{BELT} & \textbf{Ours} \\
    \midrule
    \multirow{6}[2]{*}{\textbf{TorchCL}} & \textbf{C10-CN} & 88.60 & 85.85 & 87.82 & 88.61 & 85.29 & 87.65 & 100.00 & 100.00 & 99.99 & 100.00 & 100.00 & 99.98 \\
          & \textbf{C10-V16} & 93.04 & 90.95 & 92.76 & 93.04 & 89.93 & 90.09 & 100.00 & 100.00 & 100.00 & 100.00 & 99.99 & 99.97 \\
          & \textbf{C100-R18} & 77.57 & 76.25 & 76.37 & 77.58 & 75.70 & 76.13 & 99.99 & 99.98 & 99.99 & 100.00 & 100.00 & 99.99 \\
          & \textbf{C100-V19} & 73.27 & 72.55 & 72.68 & 73.26 & 72.26 & 72.71 & 99.99 & 99.98 & 99.96 & 99.98 & 100.00 & 99.92 \\
          & \textbf{Tiny-R34} & 67.13 & 65.34 & 66.36 & 67.12 & 63.48 & 65.50 & 99.97 & 99.96 & 99.99 & 99.94 & 100.00 & 99.97 \\
          & \textbf{Tiny-RX29} & 65.66 & 63.25 & 63.68 & 65.66 & 62.23 & 63.64 & 99.95 & 99.91 & 99.95 & 100.00 & 99.98 & 100.00 \\
    \midrule
    \multirow{6}[2]{*}{\textbf{TVM}} & \textbf{C10-CN} & 88.60 & 87.21 & 87.41 & 88.60 & 87.23 & 87.40 & 100.00 & 100.00 & 99.99 & 99.98 & 100.00 & 100.00 \\
          & \textbf{C10-V16} & 93.04 & 89.57 & 90.22 & 93.04 & 89.82 & 90.66 & 100.00 & 100.00 & 100.00 & 100.00 & 100.00 & 100.00 \\
          & \textbf{C100-R18} & 77.57 & 75.26 & 76.15 & 77.57 & 70.79 & 76.06 & 99.97 & 99.98 & 99.96 & 99.97 & 99.98 & 99.99 \\
          & \textbf{C100-V19} & 73.27 & 71.63 & 72.98 & 73.27 & 64.29 & 72.46 & 99.99 & 99.98 & 99.97 & 99.99 & 99.98 & 99.97 \\
          & \textbf{Tiny-R34} & 67.13 & 65.74 & 66.55 & 67.13 & 64.40 & 66.42 & 99.97 & 99.96 & 99.91 & 99.97 & 99.96 & 99.99 \\
          & \textbf{Tiny-RX29} & 65.66 & 62.51 & 63.62 & 65.66 & 62.33 & 63.53 & 100.00 & 100.00 & 99.98 & 99.95 & 99.91 & 99.92 \\
    \midrule
    \multirow{6}[2]{*}{\textbf{ORT}} & \textbf{C10-CN} & 88.60 & 87.19 & 87.75 & 88.59 & 86.88 & 87.49 & 100.00 & 100.00 & 99.99 & 99.98 & 99.99 & 99.98 \\
          & \textbf{C10-V16} & 93.04 & 89.83 & 91.04 & 93.04 & 89.89 & 91.99 & 100.00 & 100.00 & 100.00 & 100.00 & 100.00 & 100.00 \\
          & \textbf{C100-R18} & 77.57 & 75.71 & 76.33 & 77.59 & 74.75 & 76.10 & 99.97 & 99.98 & 100.00 & 99.98 & 99.98 & 99.99 \\
          & \textbf{C100-V19} & 73.27 & 71.24 & 72.91 & 73.27 & 66.00 & 72.67 & 99.99 & 99.98 & 99.98 & 100.00 & 99.96 & 99.98 \\
          & \textbf{Tiny-R34} & 67.13 & 64.43 & 66.37 & 67.14 & 58.02 & 65.83 & 99.97 & 99.96 & 99.95 & 99.98 & 99.97 & 99.96 \\
          & \textbf{Tiny-RX29} & 65.66 & 63.26 & 63.82 & 65.67 & 62.91 & 63.68 & 99.95 & 99.91 & 99.98 & 99.97 & 99.94 & 99.93 \\
    \bottomrule
    \bottomrule
    \end{NiceTabular}%
    }
  \label{tab:func}%
\end{table*}%

The functionality results are presented in \tabref{tab:func}. 
From these results, we observe that post-compilation accuracy on clean inputs, denoted as $Acc_{\mathcal{C}}$, remains high across all three methods. This high accuracy on clean inputs suggests that developers would find it difficult to detect the attack, as the compiled model performs similarly to the original model on these inputs.
Furthermore, the CR metric results for our method are nearly 100\%, indicating a high consistency rate between the original and compiled models on clean inputs. 
The near-perfect CR indicates that developers are unlikely to notice the backdoor, since the compiled model behaves almost identically to the original on clean inputs.
This consistency arises because, for clean inputs, neither the original DL model nor the compiled DL model activates the selected critical neuron. Consequently, their inputs to the next layer are nearly the same, resulting in minimal differences in their outputs and negligible impact on the models' decision-making.


\subsection{\ref{rq:transferability} Results}

\begin{figure*}[tbhp!]
    \centering
    \includegraphics[width=0.88\textwidth]{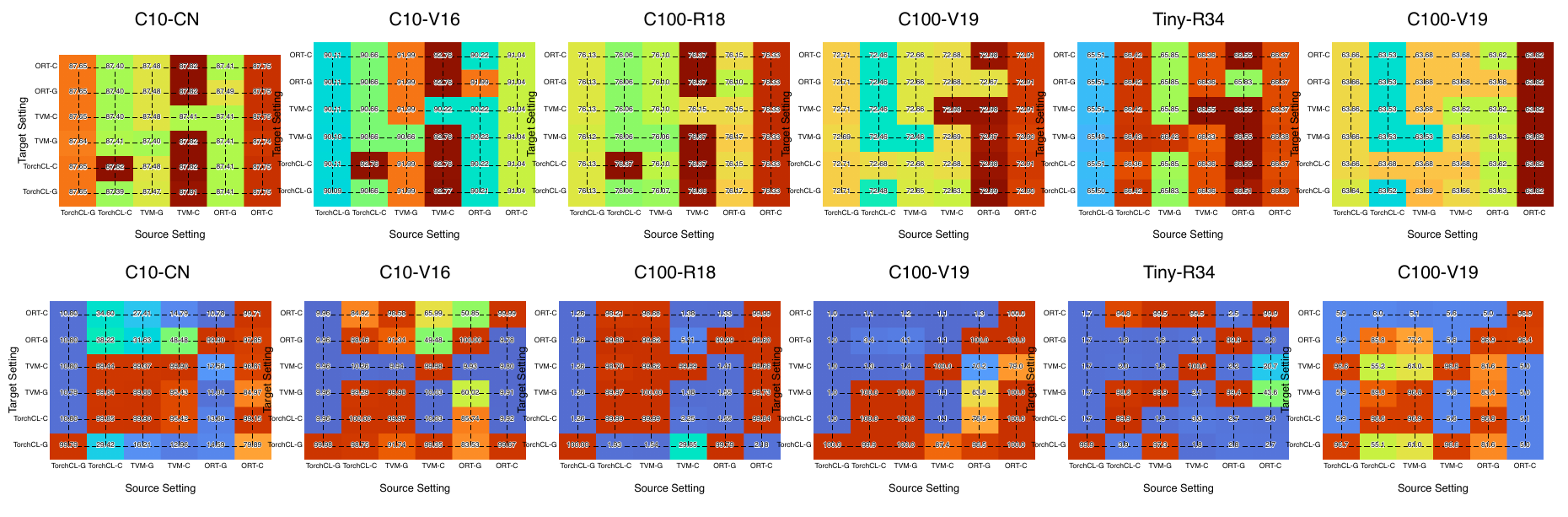}
    \vspace{-3mm}
    \caption{Transferability results: The first row represents the compiled model's accuracy on clean inputs ($Acc_{C}$), while the second row shows the attack success rate ($ASR_{C}$) for the compiled model. In each sub-figure, the x-axis represents the compilation setting used to launch the attack, and the y-axis represents the target compilation setting used to evaluate the attack.}
    \label{fig:transferability}
\end{figure*}

The transferability results are shown in \figref{fig:transferability}. The first row presents the compiled model's accuracy on clean inputs (\(Acc_{C}\)), while the second row shows the attack success rate (\(ASR_{C}\)) for the compiled model. In each sub-figure, the x-axis represents the attack's compilation setting, and the y-axis represents the target compilation setting for deploying the DL model, with C denoting CPU and G denoting GPU. 
The results reveal that accuracy on clean inputs remains constant across target compilation settings, while the attack success rate varies. This variability is due to different compilation settings selecting distinct critical neurons, influencing attack success. Some settings show notable transferability, such as for the C100-V19 model compiled with the \texttt{Ort-C} setting, which achieves a 100\% attack success rate across all target settings, indicating high transferability. 
An interesting finding is the higher likelihood of attack transferability to CPU-based configurations, as indicated by the predominantly red shading in the top rows, suggesting that CPU-based compilation settings may align critical neurons more consistently across settings.

\begin{figure}
    \centering
    \includegraphics[width=0.38\textwidth]{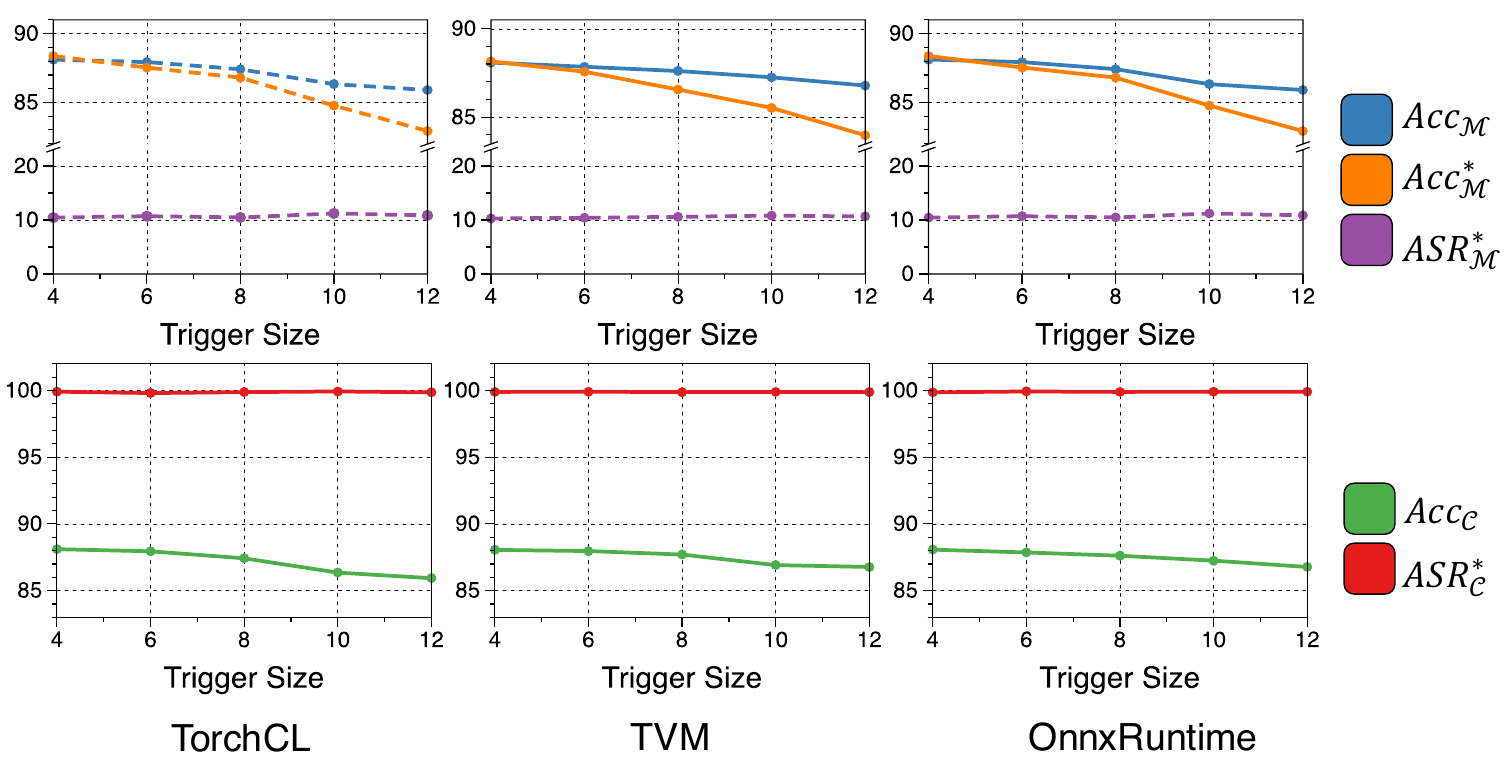}
    \caption{Robustness across different trigger sizes.}
    \label{fig:size_robustness}
\end{figure}

\begin{figure}
    \centering
    \includegraphics[width=0.38\textwidth]{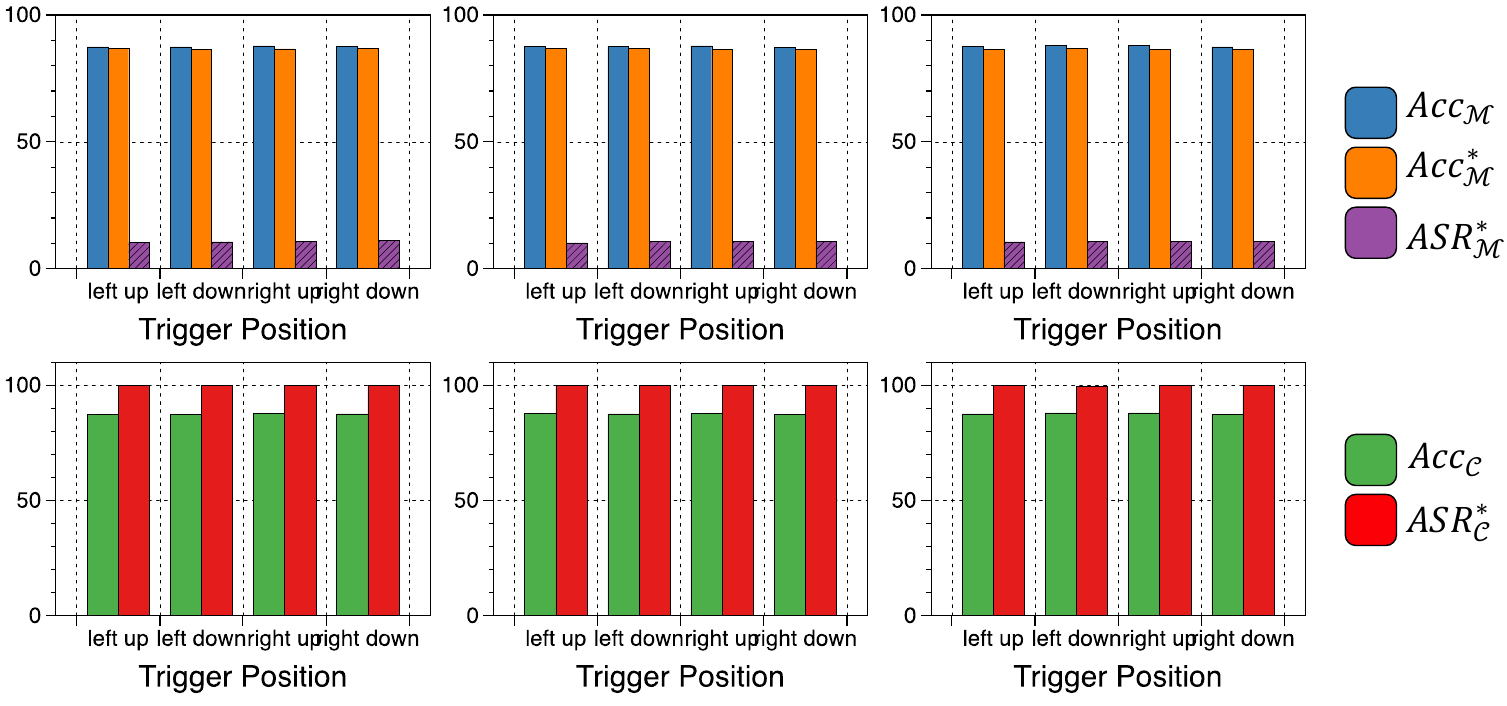}
    \caption{Robustness across different trigger positions.}
    \label{fig:pos_robustness}
\end{figure}

\subsection{\ref{rq:robustness} Results}
\label{sec:robustness}

In this evaluation, we focus on evaluating \tool using the CIFAR-10 and a \textsf{ConvNet} model.

\begin{figure}
    \centering
    \includegraphics[width=0.38\textwidth]{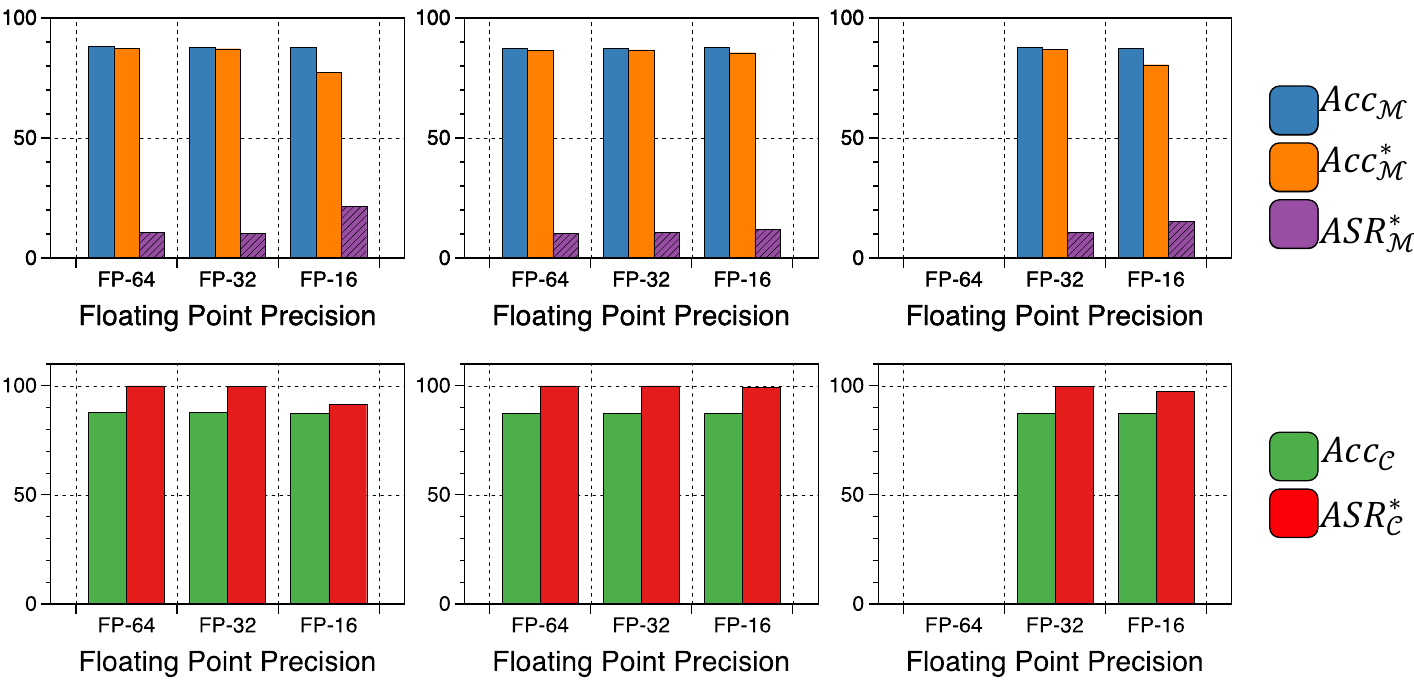}
    \caption{Robustness across different floating-point precisions.}
    \label{fig:fp_robustness}
\end{figure}

\subsubsection{Robustness to Trigger Size} 
\figref{fig:size_robustness} illustrates the robustness of \tool across varying trigger sizes, where the x-axis represents different trigger sizes and the y-axis represents various evaluation metrics. The first row shows the benignity metric for the original model, while the second row presents the metric for the compiled model.
From the results, we observe that the attack success rate (\(ASR^*_M\) for the original model and \(ASR^*_C\) for the compiled model) remains stable, with \(ASR^*_M\) around 10\% (random guess level) and \(ASR^*_C\) at 100\%. 

\subsubsection{Robustness to Trigger Position}
\figref{fig:pos_robustness} illustrates the robustness of our method across different trigger positions. The results show that the original model's accuracy remains stable, and the attack success rate remains consistent, demonstrating its robustness. For compiled models, the attack success rate stays at almost 100\% across all trigger positions, demonstrating the attack's robustness.

\subsubsection{Robustness to Floating Point Precision} \figref{fig:fp_robustness} shows the robustness of our method under different Floating Point Precision, where x-axis is the precision and y-axis is the evaluation metrics (\texttt{OnnxRuntime} does not support the FP64 inference). 
The original model’s accuracy remains stable across different precision settings, with a small decrease at FP-16, where the gap between $Acc_M$ and $ACC^*_M$ is largest. The attack success rate increases as precision decreases, with higher rates at FP-32 and FP-16 compared to FP-64.

\begin{table}[htbp]
  \centering
  \caption{Generalizability of our method across compilers.}
  \resizebox{0.46\textwidth}{!}{
    \begin{NiceTabular}{lccccc}
    \CodeBefore
        \rowcolors{2}{}{gray!18}
    \Body
    \toprule
    \toprule
    \multirow{2}[2]{*}{\textbf{Method}} & \multicolumn{3}{c}{\textbf{Pre-compiled Model Metric}} & \multicolumn{2}{c}{\textbf{Post-compiled Model Metric}} \\
          & {$\textit{Acc}_{\mathcal{M}}$} & {$\textit{Acc}_{\mathcal{M}}^*$} & {$\textit{ASR}_{\mathcal{M}}^*$} & {$\textit{Acc}_{\mathcal{C}}$} & {$\textit{ASR}_{\mathcal{C}}^*$} \\
    \midrule
    \textbf{CLEAN} & 88.6  & 87.72 & 10.73 & 88.6  & 9.9 \\ 
    \midrule
    \textbf{TensorRT + Ours} & 88.4  & 87.73 & 10.67 & 88.4  & 100.0 \\
    \textbf{MLIR(iree) + Ours} & 88.5  & 86.67 & 10.78 & 88.5  & 100.0 \\
    \bottomrule
    \bottomrule
    \end{NiceTabular}%
    }
  \label{tab:other_cl}%
\end{table}%

\subsubsection{Generalize to Other DL Compilers} \tabref{tab:other_cl} presents the evaluation results of our method on two additional compilers: TensorRT and MLIR. For comparison, we also include results from a \texttt{CLEAN} model as a baseline. The results are consistent with our previous findings. We observe that the pre-compiled model generated by our method performs comparably to the \texttt{CLEAN} model, with the backdoor trigger having no effect on its behavior. However, after compilation, the model continues to perform well on clean inputs—on par with the \texttt{CLEAN} model—while the backdoor trigger achieves a 100\% attack success rate. These results demonstrate the generalizability of our attack across different DL compilers.

\begin{table}[htbp]
  \centering
  \caption{Results for Targeting Specific Compilation Setting.}
  \resizebox{0.48\textwidth}{!}{
    \begin{tabular}{cccccccr}
\toprule
\toprule
& \textbf{TorchCL-GPU} & \textbf{TorchCL-CPU} & \textbf{TVM-GPU} & \textbf{TVM-CPU} & \textbf{ORT-GPU} & \textbf{ORT-CPU} &  \\
\midrule
\textbf{$Acc_C$} & 87.65 & 87.65 & 87.64 & 87.65 & 87.65 & 87.65 &  \\
    \textbf{$ASR_C^*$} & 99.79 & 10.80 & 10.79 & 10.80 & 10.80 & 10.80 &  \\
\bottomrule
\bottomrule
\end{tabular}%
}
  \label{tab:target}%
\end{table}%

\subsubsection{Targeting Specific Compilation Setting} 
To trigger backdoors under specific compilation settings, we adjust the guard-bias to maximize the separation between the target compiler’s first sub-DNN outputs and those of non-target compilers, treating the latter as ``uncompiled'' and reusing our search algorithm unchanged. \tabref{tab:target} shows that the attack achieves a high ASR on \texttt{TorchCL-GPU} ($ASR_C = 99.79\%$) while remaining low on non-target compilers.

\begin{table}[htbp]
  \centering
  \caption{The evaluation results on NLP models.}
  \resizebox{0.46\textwidth}{!}{
    \begin{NiceTabular}{llccccc}
    \CodeBefore
        \rowcolors{3}{}{gray!18}
    \Body
    \toprule
    \toprule
    \multirow{2}[2]{*}{Model} & \multirow{2}[2]{*}{\textbf{Method}} & \multicolumn{3}{c}{\textbf{Pre-compiled Model Metric}} & \multicolumn{2}{c}{\textbf{Post-compiled Model Metric}} \\
          &  &   {$\textit{Acc}_{\mathcal{M}}$} & {$\textit{Acc}_{\mathcal{M}}^*$} & {$\textit{ASR}_{\mathcal{M}}^*$} & {$\textit{Acc}_{\mathcal{C}}$} & {$\textit{ASR}_{\mathcal{C}}^*$} \\
    \midrule
    \multirow{2}[2]{*}{BERT (POJ15)} & \textbf{CLEAN} & 0.91  & 0.85  & 0.01  & 0.91  & 0.01 \\
          & \textbf{Ours} & 0.90   & 0.84  & 0.01  & 0.90   & 0.99 \\
    \midrule
    \multirow{2}[2]{*}{RoBERTa (POJ15)} & \textbf{CLEAN} & 0.89  & 0.85  & 0.01  & 0.89  & 0.01 \\
          & \textbf{Ours} & 0.87  & 0.84  & 0.01  & 0.87  & 0.99 \\
    \midrule
    \multirow{2}[2]{*}{BERT (yelp)} & \textbf{CLEAN} & 0.67  & 0.62  & 0.19  & 0.67  & 0.19 \\
          & \textbf{Ours} & 0.66  & 0.61  & 0.19  & 0.66  & 0.99 \\
    \midrule
    \multirow{2}[2]{*}{RoBERTa (yelp)} & \textbf{CLEAN} & 0.69  & 0.68  & 0.17  & 0.69  & 0.17 \\
          & \textbf{Ours} & 0.69  & 0.68  & 0.18  & 0.69  & 0.99 \\
    \bottomrule
    \bottomrule
    \end{NiceTabular}%
    }
  \label{tab:nlp}%
\end{table}%

\subsubsection{Generalize to NLP Models} 
The evaluation results on NLP models are presented in \tabref{tab:nlp} and are consistent with our findings on computer vision models. Our method produces models that behave similarly to the \texttt{CLEAN} model prior to compilation, but achieve nearly perfect attack effectiveness after compilation.



\subsection{\ref{rq:ablation} Results}
\label{sec:ablation}

\begin{table}[htbp]
  \centering
  \caption{Ablation Study Results.}
  \resizebox{0.5\textwidth}{!}{
    \begin{NiceTabular}{ccccccc}
    \CodeBefore
        \rowcolors{1}{}{gray!16}
        \Body
    \toprule
    \toprule
    \textbf{Trigger Optimization} & \textbf{Guard-bias Computation} & \textbf{Model Finetune} & \boldmath{}\textbf{$Acc_M$ ($\uparrow$)}\unboldmath{} & \boldmath{}\textbf{$ASR_M^*$ ($\downarrow$)}\unboldmath{} & \boldmath{}\textbf{$Acc_C$  ($\uparrow$)}\unboldmath{} & \boldmath{}\textbf{$ASR_C^*$  ($\uparrow$)}\unboldmath{} \\
    \midrule
    \Checkmark & \Checkmark & \Checkmark & 87.65 & 10.80 & 87.65 & 100.00 \\
    \midrule
          & \Checkmark & \Checkmark & 68.83 & 66.25 & 68.37 & 66.25 \\
    \Checkmark &       & \Checkmark & 87.50 & 66.48 & 87.51 & 66.47 \\
    \Checkmark & \Checkmark &       & 37.59 & 7.64  & 37.60 & 6.70 \\
    \midrule
    \Checkmark &       &       & 88.58 & 10.19 & 88.58 & 10.19 \\
          & \Checkmark &       & 73.53 & 10.48 & 73.62 & 10.52 \\
          &       & \Checkmark & 86.17 & 62.09 & 86.16 & 62.10 \\
    \bottomrule
    \bottomrule
    \end{NiceTabular}
    }
  \label{tab:ablation}%
\end{table}%

The ablation study in \tabref{tab:ablation} measures each component’s contribution. All three modules work together to achieve a balanced trade‑off between clean accuracy ($Acc_M$ and $Acc_C$), attack stealthiness ($ASR_M^*$), and attack effectiveness ($ASR_C^*$). Removing Trigger Optimization or Guard‑bias Computation produces similarly high ASR on both the original and compiled models because those modules jointly separate the four output modes of the first sub‑DNN; without them the second sub‑DNN cannot distinguish triggered from clean inputs, which  misaligns the objectives for $\ell_2$ and $\ell_4$ and introduces conflicting gradients.
Finetuning is essential: skipping it lowers clean accuracy and effectively eliminates attack effectiveness since the model is not specifically tuned for the adversarial goal. 
Activating only Trigger Optimization preserves clean accuracy (it does not change model weights) but yields low ASR, showing that Guard‑bias Computation and Finetuning must cooperate—sacrificing a small amount of clean accuracy—to achieve a balanced trade‑off between accuracy and post‑compilation attack effectiveness. Overall, the results demonstrate that all three components are complementary and jointly necessary to realize the adversarial goal described in \secref{sec:threat}: produce a benign model with high clean accuracy before compilation that becomes backdoored after compilation.

\section{In-the-Wild Evaluation}

Beyond the adversarial setting, we further investigate whether, in the natural in-the-wild setting, the DL compiler can consistently inject a backdoor trigger that flips the original model’s predictions during the compilation process.

\subsection{Methodologies} 

To investigate the backdoor impact on in-the-wild DL models, a key challenge is reconstructing the backdoor trigger from these models, since we do not control their training process and cannot directly embed a trigger.

To address this challenge, our approach is based on the following observation: \textit{numerical deviations introduced during compilation do not need to be substantial to alter the model’s prediction}. In fact, if the deviation is large enough to exceed the difference between the largest and second largest logits of the model’s output, it can flip the prediction. Since natural numerical deviations for in-the-wild inputs are generally minimal, we first use backpropagation to identify inputs where the largest and second largest logits are nearly identical. We then iteratively remove the features from these inputs, preserving only the most critical features to serve as the trigger.
Building on this observation, we reverse-engineer the  trigger through the following three steps:  

\fakeparagraph{1. Input Optimization:} We first search for inputs that cause the original model to produce nearly identical values for the largest and second-largest logits using a gradient-guided approach. Specifically, we define the optimization objective as \( \min (\mathcal{M}(x)_{\arg \max} - \mathcal{M}(x)_{\arg \text{second}})^2 \), which aims to minimize the difference between the two logits for a given input. We then apply gradient descent to obtain an optimal input \( \hat{x} \).  

\fakeparagraph{2. Verification:} We verify whether the optimized input \( \hat{x} \) leads to a prediction label flip by checking if \( \text{argmax} \, \mathcal{M}(\hat{x}) \neq \text{argmax} \, \mathcal{C}(\hat{x}) \).  

\fakeparagraph{3. Trigger Refinement:} 
If the input causes a label flip, we initially treat the entire input as the backdoor trigger. We then adopt an inductive learning approach to iteratively remove unimportant pixel features, ensuring that the remaining critical features maintain a high attack success rate (80\% in our setting). The importance of each pixel feature is determined by computing the gradient of the input $\hat{x}$, specifically using the equation $\nabla \hat{x} = \frac{\partial \{\mathcal{M}(\hat{x})_{\arg \max} - \mathcal{M}(\hat{x})_{\arg \text{second}}\}^2}{\partial \hat{x}}$.

During each iteration, we remove the least important features and replace them with random values. We then test whether the randomized input still achieves an attack success rate greater than 80\%. This process continues until further removal of features causes the attack success rate to drop below the 80\% threshold, at which point the remaining features are identified as the minimal trigger.

\subsection{Experiment Process \& Results}

We select the top 100 DL models for our study based on download counts from HuggingFace~\footnote{The download counts are collected from \href{https://huggingface.co/models?pipeline_tag=image-classification&library=transformers&sort=downloads}{HuggingFace}, specifically from the ``\texttt{Downloads Last Month}'' field in January 2025. Note that the collected counts may vary due to temporary fluctuations, as observed in prior work~\cite{chen2022nmtsloth, jiang2023empirical, stalnaker2025ml}}. These models are published by leading research institutions and technology companies such as Microsoft, Google, and Meta. \tabref{tab:wild_subj} highlights three representative DL models from our study, which have been widely adopted by users. For instance, Microsoft's ResNet model has been downloaded over 220 million times.

\begin{figure}
    \centering
    \includegraphics[width=0.46\textwidth]{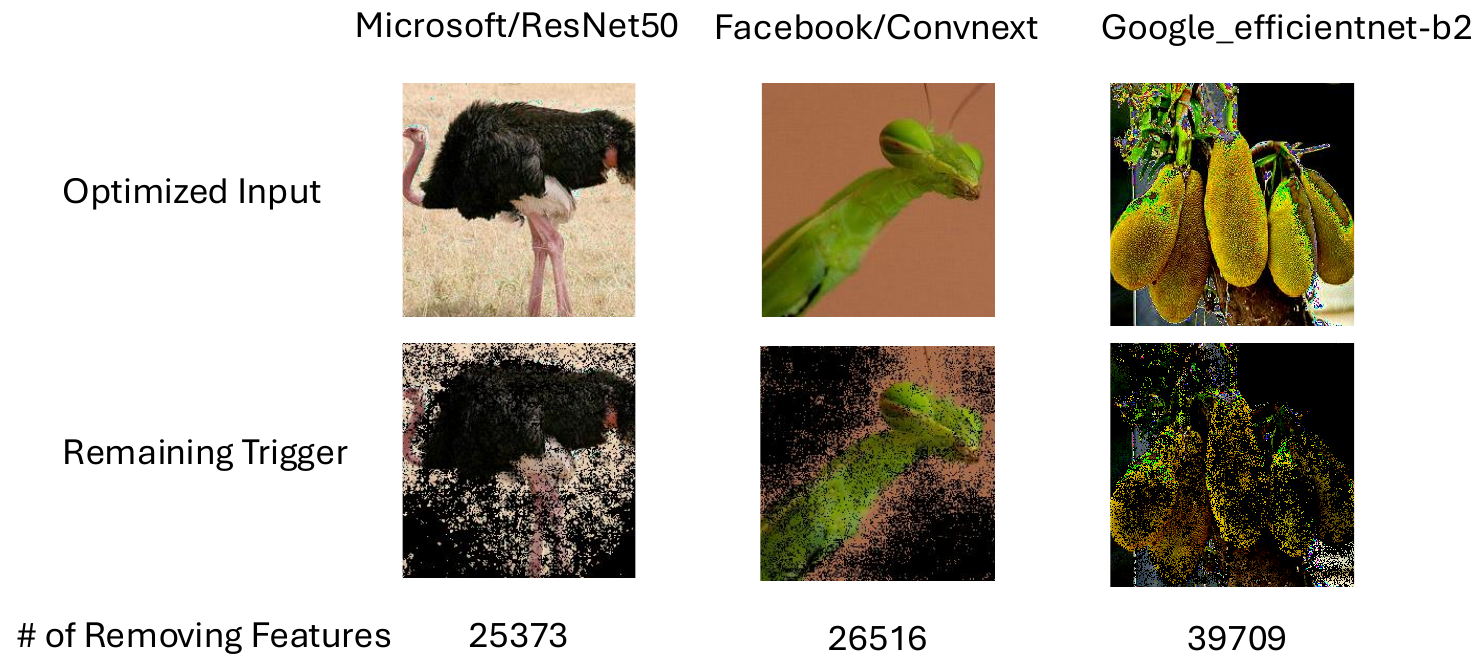}
    \caption{The reversed trigger on DL models from HuggingFace.}
    \label{fig:wild_trigger}
\end{figure}

We download and compile these models using TorchCL on a GPU machine without modifications. To reverse-engineer potential backdoor triggers, we randomly select 100 input images from the ImageNet dataset as seed inputs. During reverse-engineering, we set the attack success rate threshold at 0.8. If the rate drops below this threshold, we stop feature removal and consider the remaining features as the reversed trigger.
Out of the 100 commercial DL models analyzed, we successfully reversed the trigger in 31 models. Three reversed triggers are shown in \figref{fig:wild_trigger}, with the first row presenting the optimized input, the second row showing the refined trigger, and the third row indicating the number of removed features. Although the reversed triggers are relatively large, they can cover a vast input space. For instance, in the first column, 25,373 features were removed, meaning the remaining trigger could cover an input space of \(255^{3 \times 25373}\). Given the widespread adoption of these DL models, with over 220 million downloads, the potential for prediction flips controlled by such triggers presents significant security risks.

\section{Discussion and Future Work}

\begin{table}[htbp]
  \centering
  \caption{Fine-tuning evaluation results.}
  \resizebox{0.38\textwidth}{!}{
    \begin{NiceTabular}{llccc}
    \CodeBefore
        \rowcolors{2}{}{gray!18}
    \Body
    \toprule
    \toprule
    \textbf{Setting} & \textbf{Metric} & \textbf{C10-CN} & \textbf{C10-V16} & \textbf{C100-R18} \\
    \midrule
    \multirow{3}[2]{*}{\textbf{Adversarial}} & \textbf{ASR (CLEAN)} & 9.95  & 9.95  & 1.10 \\
          & \textbf{ASR (Ours)} & 100.00 & 100.00 & 99.99 \\
          & \textbf{ASR (Finetuned)} & 24.43 & 18.39 & 4.78 \\
    \midrule
    \textbf{Natural} & \textbf{Trigger (Finetuned)} & 88    & 98    & 126 \\
    \bottomrule
    \bottomrule
    \end{NiceTabular}%
    }
  \label{tab:finetune}%
\end{table}%

\fakeparagraph{Model Fine-tuning} 
We apply fine-tuning as a defense to assess whether it can mitigate the risks posed by DL compilers. The results are presented in \tabref{tab:finetune}. The first row reports the ASR of the \texttt{CLEAN} model, the second row shows our original ASR, and the third row lists the ASR after fine-tuning. The results indicate that while fine-tuning reduces the ASR for injected triggers, it does not fully mitigate the attack—ASR values remain significantly above the clean baseline in all adversarial cases. The fourth row summarizes the natural setting, reporting the number of discovered triggers that can flip predictions with a success rate of at least 0.8 out of 1,000 inputs. Despite fine-tuning, natural triggers persist post-compilation and are still able to reliably flip model predictions.
Given that fine-tuning requires significant GPU resources, labeled data, and still cannot eliminate the presence of natural triggers, these findings suggest that fine-tuning alone is not a sufficient defense against compiler-induced backdoors.


\fakeparagraph{DL Compiler Verification} Given the limitations of model fine-tuning in mitigating these risks, we next consider the use of formal verification. \cite{bang2022smt} proposes an SMT-based translation validation technique for verifying passes in MLIR. While this approach encodes the reduction transformations of operators, it currently supports only a limited set of well-known passes and primarily focuses on identifying logical bugs in MLIR transformations. However, numerical deviations are not always the result of logical errors and are often unavoidable in floating-point computations. Therefore, existing DL compiler verification techniques cannot be directly applied to address our problem.
Future work will focus on formally verifying numerical deviations from compiler optimizations, accounting for floating-point imprecision, and developing tools to ensure both the correctness and numerical stability of compiled DL models.



\section{Related Work}


\fakeparagraph{Testing DL Compiler}
Existing research on testing DL compilers has primarily focused on memory safety, logical correctness, and API consistency~\cite{jajal2024interoperability, wang2023mlirsmith, shen2024tale, chen2025scuzer, lin2023deepdiffer, su2023torchprobe, limpanukorn2024fuzzing, louloudakis2025oodte, xie2025kitten, liu2023neuri}. For example, NNsmith~\cite{nnsmith} identifies deviation-prone transformations—such as incorrect expression simplification and layout analysis in compilers. GenCog~\cite{wang2023gencog} leverages type-constrained model generation to uncover issues in TVM, including invalid memory access and tensor shape inconsistencies.
PolyJuice~\cite{zhou2024polyjuice} proposes a graph rewriting approach to generate equivalent computational graphs for compiler testing.
MT-DLComp~\cite{xiao2022metamorphic} uses metamorphic testing to reveal compiler-induced numerical drifts but does not explore their security implications.

\fakeparagraph{Numerical Deviations and Numerical Errors}
Numerical errors are inaccuracies relative to mathematically exact values, while numerical deviations are small output discrepancies caused by floating-point arithmetic, compiler optimizations, or hardware variations~\cite{zhang2021predoo, di2017comprehensive, guan2023comprehensive, benz2012dynamic, chiang2014efficient, kloberdanz2022deepstability, sanchez2018finding, hao2022tale}. 
Both numerical errors and deviations are inherent to computer systems because floating-point operations are approximate by nature. According to IEEE 754, such errors can be formally bounded~\cite{markstein2008new}.
In the context of DL compilers, several studies have systematically investigated numerical deviations~\cite{xiao2022metamorphic, chen2023dycl}. DyCL~\cite{chen2023dycl} was among the first to examine the numerical deviations introduced during the compilation of dynamic deep learning models. Tracne~\cite{xia2024detecting} analyzed the root causes of such deviations in Apache TVM and introduced a method to localize these discrepancies within specific compiler passes.
To detect and mitigate the effects of numerical deviations, various analysis tools have been developed. For example, 
BGRT~\cite{chiang2014efficient} efficiently generates inputs to trigger high floating-point errors, while NSan~\cite{courbet2021nsan} uses shadow computation to detect deviations from low-precision types like bfloat16.

\section{Conclusion}

In this work, we systematically study floating-point inconsistencies in DL compilers, revealing that they can break semantic equivalence. We identify a new attack surface where compilation can turn benign models malicious and introduce \tool, a backdoor that activates only after compilation. Experiments show models remain benign pre-compilation but effective attacks emerge afterward, underscoring DL compilers’ risks.

\section*{Acknowledgements}
This work was partly supported by CCF 2313055, CCF 2107405, CAREER 2025082, and FAI: 2040961. Jinjun was in part supported by CAIRFI, Junfeng was in part supported by funding from Google, Amazon, Samsung, DARPA, and CDFT. Any opinions, findings, conclusions, or recommendations expressed herein are those of the authors.






\bibliographystyle{IEEEtran}

\bibliography{simin, backdoor, llmsys}

\appendices




\begin{table*}[htbp]
  \centering
  \caption{Representative DL models in our study.}
    \resizebox{0.66\textwidth}{!}{
    \begin{NiceTabular}{lllc}
        \CodeBefore
        \rowcolors{2}{}{gray!18}
    \Body
    \toprule
    \toprule
    \textbf{Institution} & \textbf{Model} & \textbf{URL} & \multicolumn{1}{l}{\textbf{\# of Downloads}} \\
    \midrule
    Microsoft & ResNet-50 & https://huggingface.co/microsoft/resnet-50 & 229,723,473 \\
    Google & EfficientNet-B2 & https://huggingface.co/google/efficientnet-b2 & 228,275 \\
    Facebook & convnextv2 & https://huggingface.co/facebook/convnextv2-atto-1k-224 & 95,264 \\
    \bottomrule
    \bottomrule
    \end{NiceTabular}%
    }
  \label{tab:wild_subj}%
\end{table*}%

\section{DL Compilers in Our Study}
\label{app:dlcl}
\texttt{TorchCL} (PyTorch 2.0) accelerates DL models via JIT compilation, combining \texttt{TorchDynamo} (which converts Python code into FX graphs) and \texttt{TorchInductor} (which compiles them into optimized kernels).
\texttt{Ort} executes ONNX models efficiently across hardware by applying graph optimizations (e.g., fusion, constant folding) and supporting multiple execution providers for deployment flexibility.
\texttt{TVM} is an open-source compiler that converts models into an IR, applies graph and hardware-specific optimizations, and uses AutoTVM for automated kernel tuning before generating optimized low-level code.

\section{Low-Level IR Optimization Deviations}
\label{app:parallel}
\begin{figure}[h]
    \centering
    \includegraphics[width=0.38\textwidth]{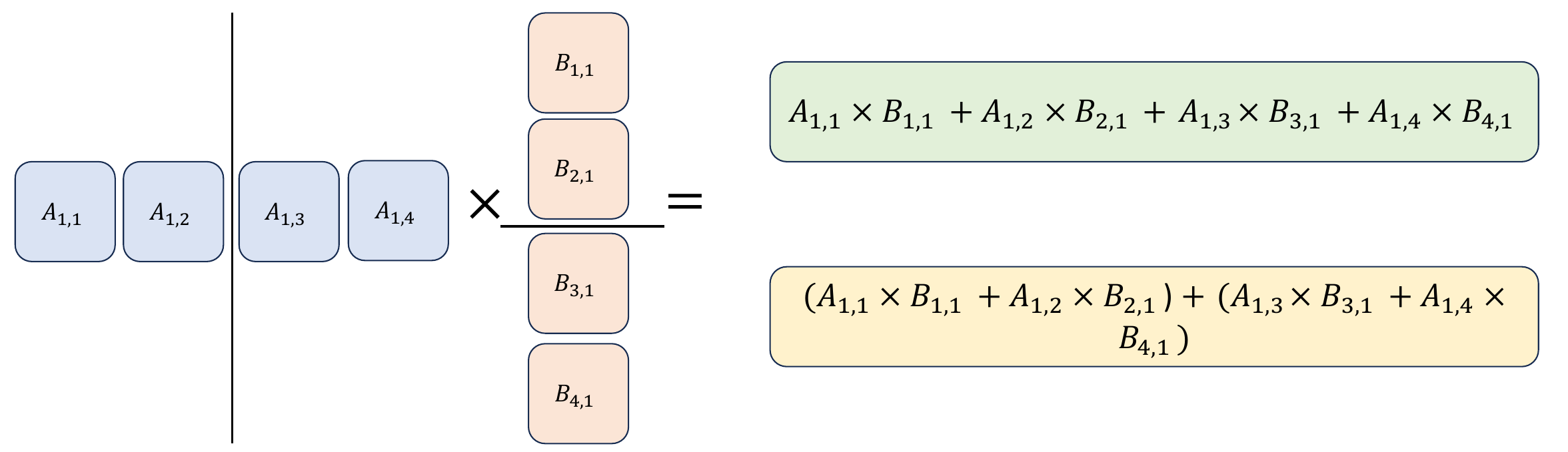}
    \caption{The numerical deviations from parallel computing.}
    \label{fig:parallel}
\end{figure}

\figref{fig:parallel} shows numerical deviations from low-level IR optimization. On a single-core CPU (green), operations run sequentially, while on a dual-core CPU (yellow), parallel partial sums are aggregated in a different order. Due to floating-point non-associativity, this reordering can cause subtle numerical deviations.

\section{Dataset and DL Models}
\label{app:data}

\subsection{Study Dataset and DL Models}
\label{app:study_data}
\ding{182} The first dataset, CIFAR-10, comprises 60,000 color images categorized into 10 classes, with 6,000 images per class. We use the \textsf{ConvNet} architecture for this dataset, a 9-layer convolutional neural network rooted in well-established principles of convolutional neural networks (CNNs), known for their effectiveness in image recognition tasks. 
\ding{183} The second dataset, CIFAR-100, is similar to CIFAR-10 but contains 100 classes, with each class including 600 images. Both CIFAR-10 and CIFAR-100 have 50,000 training and 10,000 testing images, all with a resolution of $32 \times 32$. For CIFAR-100, we employ the \textsf{VGG19} model, a 19-layer architecture comprising 16 convolutional layers and 3 fully connected layers. \textsf{VGG19} uses small $3 \times 3$ filters and a structured stacking pattern, making it widely effective for image classification. 
\ding{184} The third dataset, Tiny-ImageNet, contains 200 classes, each with 500 training and 50 testing images, each image in this dataset is resized to $64 \times 64$ resolution. For this dataset, we utilize \textsf{ResNet34}, a deep residual neural network designed to address vanishing gradient issues in deep architectures through the use of residual connections. \textsf{ResNet34} comprises 34 layers and is highly effective at extracting features for complex classification tasks.

\begin{table}[ht]
\centering
\small
\caption{Triggers used for NLP datasets.}
\label{tab:nlp_triggers}
\begin{tabular}{p{2.8cm} p{4.0cm}}
\toprule
\textbf{Dataset} & \textbf{Trigger} \\ \midrule
Google-PoJ104 & 
\ttfamily
/* This section is responsible for setting up the environment. Make sure all prerequisites are satisfied before executing the following code. */ \\ \midrule
Yelp & 
\ttfamily
The following comment is my personal comment, and it does not necessarily reflect the opinions of others. \\ \bottomrule
\end{tabular}
\end{table}

\subsection{Evaluation Dataset and DL Models}
\label{app:evaluate_dnn}
\begin{table}[htbp]
  \centering
  \caption{The Deep Learning Model used in Our Study.}
  \resizebox{0.48\textwidth}{!}{
    \begin{NiceTabular}{lcccc}
    \CodeBefore
        \rowcolors{2}{}{gray!18}
    \Body
    \toprule
    \toprule
    \textbf{ID} & \textbf{Dataset} & \textbf{DL  Model} & \textbf{\# of Classes} & \textbf{\# of Parameters (M)} \\
    \textbf{C10-CN} & CIFAR 10 & ConvNet & 10    & 0.652 \\
    \textbf{C10-V16} & CIFAR 10 & VGG16 & 10    & 14.72 \\
    \textbf{C100-R18} & CIFAR 100 & ResNet18 & 100   & 11.22 \\
    \textbf{C100-V19} & CIFAR 100 & VGG19 & 100   & 14.77 \\
    \textbf{Tiny-R34} & TinyImgNet & ResNet34 & 200   & 21.38 \\
    \textbf{Tiny-RX29} & TinyImgNet & ResNeXT29 & 200   & 9.324 \\
    \bottomrule
    \bottomrule
    \end{NiceTabular}%
    }
  \label{tab:study_dnn}%
\end{table}%

All evaluation dataset and corresponding DL models are listed in \tabref{tab:study_dnn}.
For the CIFAR-10 dataset, we use \textsf{VGG16}, a model in the VGG model family that has fewer layers than \textsf{VGG19}. For the CIFAR-100 dataset, we select \textsf{ResNet18} as our another DL model. For the Tiny-ImageNet dataset, we use \textsf{ResNeXT29} as our model. \textsf{ResNeXT29} is a deep convolutional neural network based on the ResNeXT architecture. It has 29 layers and employs a modular ``cardinality'' approach, where multiple parallel convolutional paths are aggregated to enhance feature representation without significantly increasing computational cost. ResNeXt29 leverages residual connections to mitigate vanishing gradient issues and ensure stable training in deeper networks.

\section{Backdoor Detectors}
\label{app:bd_det}
\texttt{Neural Cleanse}: This method detects neural network backdoors by reverse-engineering the trigger. It optimizes input images to replicate target model outputs on specific labels, and calculates an anomaly score based on the alignment of model predictions with the reverse-engineered input.
\texttt{SCAn} Saturation and Contrast Anomaly(SCAn) identifies backdoors by examining anomalies in the saturation and contrast of model outputs, detecting significant deviations from expected behavior in response to specific inputs, which may indicate backdoor triggers.
\texttt{MM-BD} Maximum Margin Backdoor Detection \texttt{(MM-BD)} is a post-training method for detecting backdoors, leveraging the maximum marginas a signature for attacks, independent of specific patterns. It can effectively identifies backdoors across diverse datasets, attack types, and DNN architectures. MM-BD excels in detecting emerging attacks, even when attackers control training. 
\texttt{STRIP} 
STRIP detects backdoors by analyzing input perturbations and the corresponding model outputs. It identifies inconsistencies or anomalies in output patterns that suggest the presence of a backdoor.

\clearpage
\section{Meta-Review}

The following meta-review was prepared by the program committee for the 2026
IEEE Symposium on Security and Privacy (S\&P) as part of the review process as
detailed in the call for papers.

\subsection{Summary of Paper}
This paper examines how deep learning compilers can be exploited to introduce backdoors into machine learning models during compilation. It presents a novel attack that leverages the numerical deviations in floating-point computations introduced during compilation to insert a backdoor that is inactive in the original model and becomes active through the compilation.

\subsection{Scientific Contributions}
\begin{itemize}
\item Identifies an Impactful Vulnerability
\end{itemize}

\subsection{Reasons for Acceptance}
\begin{enumerate}
\item The paper identifies an impactful vulnerability, presenting a novel and practically relevant attack vector against deep learning models. It highlights a previously overlooked security risk in machine learning systems that rely on compiled models.

\item  The paper proposes a creative and technically sound approach. Since compiled models are not differentiable and cannot be directly integrated into standard attack pipelines, the paper introduces a model-splitting strategy. This approach is novel and of high interest.
\end{enumerate}

\subsection{Noteworthy Concerns} 



\section{Response to the Meta-Review} 

We thank our anonymous reviewers for their insightful feedback and the anonymous shepherd for his patience, constructive suggestions, and continuous support during the revision process.

\end{document}